\documentclass[sigconf]{acmart}
\usepackage{multirow} 
\usepackage{makecell}
\usepackage{algorithm}
\usepackage{algpseudocode}
\usepackage{wasysym}
\usepackage{subcaption} 
\usepackage{xcolor}  
\usepackage{colortbl}
\usepackage[table]{xcolor}  
\AtBeginDocument{%
  }

\setcopyright{acmlicensed}
\copyrightyear{2018}
\acmYear{2018}
\acmDOI{XXXXXXX.XXXXXXX}
\acmConference[Conference acronym 'XX]{Make sure to enter the correct
  conference title from your rights confirmation email}{June 03--05,
  2018}{Woodstock, NY}
\acmISBN{978-1-4503-XXXX-X/2018/06}

\begin{document}

\title{When Reasoning Leaks Membership: Membership Inference Attack on Black-box Large Reasoning Models}


\author{Ruihan Hu}
\affiliation{%
  \institution{Beijing University of Posts and Telecommunications}
  \city{Beijing}
  \country{China}}
\email{gloria-1019@bupt.edu.cn}

\author{Yu-Ming Shang}
\affiliation{%
  \institution{Beijing University of Posts and Telecommunications}
  \city{Beijing}
  \country{China}}
\email{	shangym@bupt.edu.cn}

\author{Wei Luo}
\affiliation{%
  \institution{Beijing University of Posts and Telecommunications}
  \city{Beijing}
  \country{China}}
  \email{luowei@bupt.edu.cn}

\author{Ye Tao}
\affiliation{%
 \institution{China Unicom Research Institute}
  \city{Beijing}
  \country{China}}
 \email{taoy10@chinaunicom.cn}

\author{Xi Zhang}
\authornote{Corresponding author.}
\affiliation{%
  \institution{Beijing University of Posts and Telecommunications}
  \city{Beijing}
  \country{China}}
\email{zhangx@bupt.edu.cn}

\renewcommand{\shortauthors}{Trovato et al.}

\begin{abstract}
Large Reasoning Models (LRMs) have rapidly gained prominence for their strong performance in solving complex tasks. Many modern black-box LRMs expose the intermediate reasoning traces through APIs to improve transparency (\emph{e.g.}, Gemini-2.5 and Claude-sonnet). Despite their benefits, we find that these traces can leak membership signals, creating a new privacy threat even without access to token logits used in prior attacks. In this work, we initiate \textit{the first systematic exploration of Membership Inference Attacks (MIAs) on black-box LRMs}. Our preliminary analysis shows that LRMs produce confident, recall-like reasoning traces on familiar training member samples but more hesitant, inference-like reasoning traces on non-members. The representations of these traces are continuously distributed in the semantic latent space, spanning from familiar to unfamiliar samples. Building on this observation, we propose \textbf{BlackSpectrum}, the first membership inference attack framework targeting the black-box LRMs. The key idea is to construct a recall–inference axis in the semantic latent space, based on representations derived from the exposed traces. By locating where a query sample falls along this axis, the attacker can obtain a membership score and predict how likely it is to be a member of the training data. Additionally, to address the limitations of outdated datasets unsuited to modern LRMs, we provide two new datasets to support future research, arXivReasoning and BookReasoning. Empirically, exposing reasoning traces greatly increases the vulnerability of LRMs to MIAs, boosting attack accuracy by up to 23.8\%, AUC by 29.9\%, and nearly doubling TPR@5\%FPR. Our findings highlight the need for LRM companies to balance transparency in intermediate reasoning traces with privacy preservation.\footnote{Our code is available at \url{https://github.com/STAIR-BUPT/LRM-MIA}.}

\end{abstract}





\received{20 February 2007}
\received[revised]{12 March 2009}
\received[accepted]{5 June 2009}

\settopmatter{printacmref=false} 
\renewcommand\footnotetextcopyrightpermission[1]{} 
\pagestyle{plain} 

\maketitle

\section{Introduction}
Large Reasoning Models (LRMs) have rapidly gained prominence for their strong performance in solving complex tasks~\cite{deepseekai2025deepseekr1incentivizingreasoningcapability,kumar2025llmposttrainingdeepdive,muennighoff2025s1simpletesttimescaling}. 
Nowadays, many LRM companies would like to provide detailed reasoning traces via APIs, which can allow users to inspect intermediate reasoning steps before the final answer, even though their models remain black-box (\emph{e.g.}, Gemini-2.5~\cite{comanici2025gemini25pushingfrontier} and Claude-sonnet~\cite{anthropic2025claude4}). 
This design aims to improve the interpretability and transparency of the model, helping users understand how it reasons through complex problems~\cite{chen2025policyframeworkstransparentchainofthought,bilal2025llmsexplainableaicomprehensive}.


However, while such transparency is beneficial, it also provides a more informative view into the model's generation process, potentially introducing additional security and privacy risks. In particular, the exposed reasoning traces probably leak signals about the model's training data, a classical privacy threat known as Membership Inference Attacks (MIAs)~\cite{shokri2017membershipinferenceattacksmachine,carlini2022membershipinferenceattacksprinciples}. Formally, MIA refers to the task of determining whether a specific data sample (\emph{i.e.}, query sequence in this work) is a member included in a target model’s training set, which may contain industry-confidential information~\cite{codeleak,carlini2021extracting} or copyrighted materials~\cite{cooper2025extractingmemorizedpiecescopyrighted, grynbaum2023nyt,duarte2024decop}.

%
%
 
\begin{figure}[t] 
  \includegraphics[width=1\columnwidth]{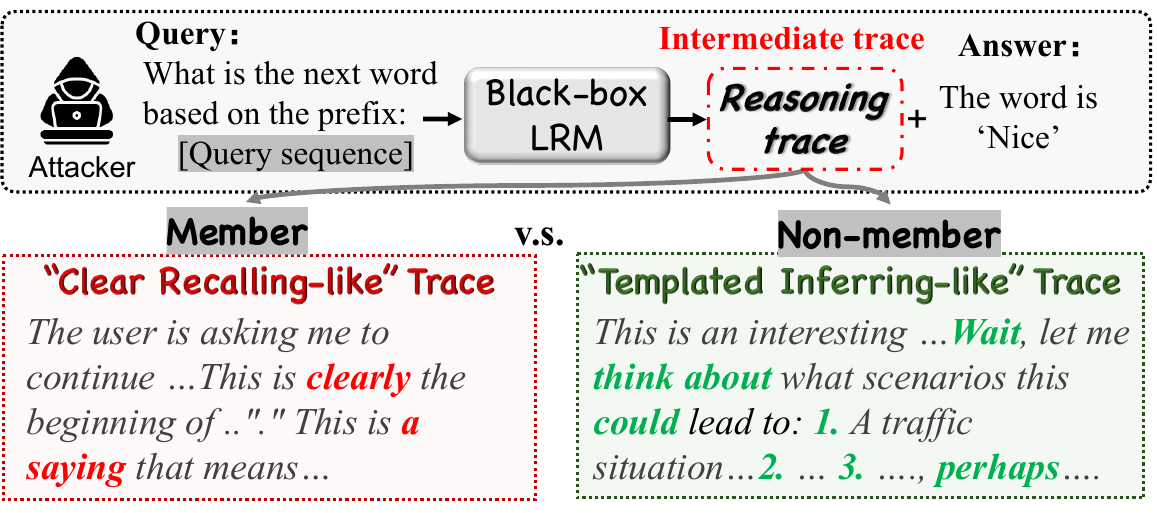}
    \Description{intro. figure}
  \caption {The real-world cases on Claude-sonnet-4~\cite{anthropic2025claude4}, the intermediate reasoning traces disclosed by the LRM's API reveal membership cues. The member trace exhibits a certain recall-like reasoning mode, whereas the non-member trace exhibits an uncertain inferential reasoning mode.}
  \label{fig: intro}
\end{figure}

In this paper, we initiate a systematic exploration of the following question in LRMs: \textit{Do reasoning traces contain membership signals that make the attack easier?} 
 As shown in Figure~\ref{fig: intro}, we begin with an illustrative example using Claude-sonnet-4~\cite{anthropic2025claude4}. The attacker prompts the target LRM with a query sequence to predict the next word. Member sequence produces reasoning traces that exhibit a clear confidence and recall-like reasoning mode, whereas non-member sequence exhibits an uncertain and template-like reasoning mode.
 This observation naturally suggests that differences in certainty could serve as heuristic signals of membership. However, turning this intuition into an effective method is non-trivial due to the following challenges. First, surface-level cues such as those based on the certainty or hesitation words (\emph{e.g.}, \textit{`clearly'} or \textit{`wait'}) are unstable (See our experimental results in Sec.~\ref{sec:explore}). This is because they are heavily influenced by context and the reasoning template styles of different target LRMs. Second, it is challenging to quantitatively assess the target LRM’s fitting degree with a sequence (\emph{i.e.}, a higher fitting degree means the model is more familiar with the sequence and thus more likely to be a member~\cite{shokri2017membershipinferenceattacksmachine}). Traditional attacks typically quantify this fitting degree using prediction confidence~\cite{zhang2025mink,he2025labelonlymembershipinferenceattack}, which is unavailable in black-box LRMs.
 

To address the above challenges, we move beyond surface-level cues and encode these reasoning traces into a latent semantic space to capture deeper patterns. In this latent space, the target LRM's fitting behavior still cannot be characterized, so we start exploring from two extreme cases: (1) highly familiar sequences whose continuations the LRM can reproduce verbatim, and (2) unfamiliar synthetic sequences that require the LRM to rely purely on its generalization ability to infer the continuation rather than memorization. 
	Interestingly, traces from these two extremes are well separated in the latent space, forming a clear direction (as illustrated in Figure~\ref{fig:pre-claude}).
Furthermore, reasoning traces of general member and non-member sequences lie between these two ends, gradually shifting from recall-like to inference-like reasoning modes as familiarity decreases. We term this phenomenon \textbf{Recall–Inference Spectrum}, describing a gradual transition of reasoning modes as sequence familiarity decreases.

Building on this observation, we propose \textbf{BlackSpectrum}, a novel MIA framework that leverages intermediate reasoning traces to target modern black-box LRMs. The core idea is to construct a recall–inference axis in the latent space from reasoning trace representations, along which the position of a query sequence reflects its likelihood of membership. BlackSpectrum consists of three components: the reasoning trace encoder, the recall-inference axis builder, and the projection-based membership predictor. The encoder transforms the query sequences' corresponding reasoning traces into representations in the latent space. The axis builder constructs a recall–inference axis within this space using two extreme cases. Finally, the predictor projects query sequences onto this axis and computes membership scores based on their relative positions.


\vspace{0.3em}
\noindent\textbf{Evaluations and Findings}. To address the limitations of outdated evaluation datasets that are unsuited to modern LRMs, we introduce two new datasets, arXivReasoning and BookReasoning, containing 184 books with 6112 sequences and 55 papers with 1065 sequences, respectively (see details in Sec.~\ref{sec: datasets}). Furthermore, we provide 4 naive reasoning trace-based attack baselines targeting black-box LRMs (see details in Sec.~\ref{sec:explore}). Evaluations on real-world LRMs show that the exposed reasoning traces contain distinctive membership signals. Our proposed framework, BlackSpectrum, achieves improvements of up to 23.8\% in accuracy, 29.9\% in AUC, and 1.9$\times$ in TPR@5\% FPR over existing attacks, underscoring the severe privacy risks introduced by exposed reasoning traces. These findings highlight the critical need to balance reasoning transparency with privacy preservation in modern LRMs.



\vspace{0.5em} 
\noindent\textbf{Contributions}. We summarize our main contributions as follows:
\vspace{-0.5em}
\begin{itemize}

   \item  To the best of our knowledge, we are the first to explore the membership inference attacks in the black-box LRMs, showing that their API-exposed reasoning traces contain	distinct membership signals. 

    \item We reveal an interesting phenomenon emerging from LRMs’ generated reasoning traces: When these traces are encoded into a latent semantic space, they form a gradual distribution from recall-like to inference-like reasoning modes as familiarity decreases. We term this observation the Recall-Inference Spectrum phenomenon.
    
    \item We propose BlackSpectrum, a new MIA framework that leverages intermediate reasoning traces against black-box LRMs. It constructs a recall–inference axis in the latent semantic space derived from reasoning trace representations, enabling attackers to infer membership by locating the query sequence along this axis.

     \item We introduce two new datasets, \textit{arXivReasoning} and \textit{BookReasoning}, for future privacy studies on LRMs, and extensively evaluate our method on modern black-box LRMs.

\end{itemize}

%
%
%
%
%

\vspace{-0.5em}
\section{Background \& Related Work}

\noindent\textbf{Large Reasoning Model Security}.
Large Reasoning Models (LRMs) have recently emerged as a more advanced class of language models~\cite{deepseekai2025deepseekr1incentivizingreasoningcapability,kumar2025llmposttrainingdeepdive,muennighoff2025s1simpletesttimescaling,ouyang2022traininglanguagemodelsfollow,muennighoff2025s1simpletesttimescaling}.
While LRMs achieve remarkable reasoning performance, they also expose new security vulnerabilities~\cite{wang2025safetylargereasoningmodels}, including overthinking~\cite{kumar2025llmposttrainingdeepdive} and reasoning-based jailbreaks~\cite{yao2025mousetrapfoolinglargereasoning}.
However, their privacy risks remain largely unexplored. We are the first to study training data membership privacy in black-box LRMs, a fundamental problem in machine learning privacy~\cite{carlini2019secret,shokri2017membershipinferenceattacksmachine,carlini2022membershipinferenceattacksprinciples}.

\vspace{0.3em}
\noindent\textbf{Membership Inference Attacks}.
Membership Inference Attacks (MIAs) aim to determine whether a specific data record is a member of the target model's training set~\cite{shokri2017membershipinferenceattacksmachine}. MIAs were first introduced in computer vision~\cite{shokri2017membershipinferenceattacksmachine}. With the rapid development of LLMs, MIAs have been gradually applied to natural language processing tasks~\cite{carlini2021extracting,shi2024detecting,he2025labelonlymembershipinferenceattack,zhang2025mink,Zhang2024PositionMI, xie-etal-2024-recall,meeus2023did, puerto-etal-2025-scaling}. 
Existing MIAs focus on traditional pre-training LLMs under the next-token prediction paradigm, many of which provide logits for output tokens (\emph{e.g.}, GPT-3~\cite{brown2020languagemodelsfewshotlearners} and LLaMA~\cite{touvron2023llamaopenefficientfoundation}). Most methods~\cite{shi2024detecting,meeus2023did,zhang2025mink,zhang2024pretrainingdatadetectionlarge} use token logit statistics to derive a membership score, assuming that members typically exhibit higher logits than non-members, as the model tends to fit members better~\cite{carlini2021extracting,puerto-etal-2025-scaling}. 
Later studies on black-box LLMs developed sampling-based MIAs that repeatedly query the model with identical prefixes and analyze suffix variations, assuming that members produce more consistent suffixes~\cite{kaneko2024samplingbasedpseudolikelihoodmembershipinference,dong-etal-2024-generalization}. In addition, some document-level MIAs (e.g., copyrighted book detection~\cite{duarte2024decop,chang-etal-2023-speak}) evaluate membership by prompting LLMs with a large number of multiple-choice~\cite{golchin2025datacontaminationquiztool} or name-cloze~\cite{chang-etal-2023-speak} questions about a document, using the number of correct answers as a membership score. Different from prior works, we focus on LRMs, a new class of models that uniquely expose reasoning traces, and study how these traces provide a new perspective for membership inference. 
\vspace{-0.3em}
\section{Problem Statement}
\noindent\textbf{Threat Model}. Following prior work on membership inference attacks~\cite{zhang2025mink,shi2024detecting,duarte2024decop,carlini2022membershipinferenceattacksprinciples}, our objective is to determine whether a query sequence is a member in the training data of the target LRM.
Formally, given a query sequence $s_i$, the attacker $\mathcal{A}$ would predict the membership status $m_{s_i}$ of the sequence $s_i$ with respect to the target LRM $\Theta$, which can be formulated as follows: 
\[
\mathcal{A}(s_i\mid \Theta, \mathcal{K}) =
\begin{cases}
m_{s_i} = 1 \ (x \in D_{\theta}) &\epsilon \geq \lambda \\
m_{s_i} = 0 \ (x \notin D_{\theta}) & \epsilon < \lambda, \tag{1}
\end{cases}
\]
where $D_{\theta}$ denotes the training dataset of the target LRM, $\epsilon$ is the membership score obtained by attack algorithm, $\lambda$ is the decision threshold, and $\mathcal{K}$ represents the attack’s prior knowledge.

\noindent\textbf{Adversary's Knowledge}. Our work considers a realistic scenario targeting commercial black-box LRMs. Specifically, the attacker $\mathcal{A}$ has \textit{no access to the model's architecture, parameters, or token logits}, but can observe the \textit{reasoning trace} exposed through the API. For example, the APIs of LRMs such as Claude-sonnet-4~\cite{anthropic2025claude4} expose reasoning traces via the \texttt{reasoning\_content} field. The recently released GPT-5-mini~\cite{openai2025gpt5} also provides a summarized version of its reasoning trace through the \texttt{reasoning\_summary} field.


\section{Preliminary Exploration} \label{sec:explore}

\begin{table}[t]
\caption{Statistical testing of membership scores from our proposed naive attacks. ES denotes the effect size.}
\centering
  \renewcommand{\arraystretch}{0.7} 
\resizebox{1\linewidth}{!}{%
\begin{tabular}{c|c|cc|c}
\specialrule{1.25pt}{0pt}{0pt}
\textbf{Model}& \textbf{Attacks} & \textbf{p-value} $\downarrow$ & \textbf{ES} $\uparrow$ & \textbf{Leak?} \\ 
\midrule
\multirow{5}{*}{\makecell{Claude-\\sonnet\\-4-thinking}} 
 & Thinking Token & $\text{2.347} \times 10^{-6}$  & 0.688 & \CIRCLE \\ 
 & Compression Rate &  $\text{5.102} \times 10^{-5}$  & 0.586 &  \CIRCLE \\  
 & Tr-Consistency (Char)  &  $\text{8.530} \times 10^{-1}$  & -0.026&  \Circle\\  
 & Tr-Consistency (Token) &  $\text{8.530} \times 10^{-1}$  & 0.026 & \Circle \\  
 & LLM-based judgement&  $\text{9.406} \times 10^{-7}$  & 0.725 & \CIRCLE \\  
  \midrule
 \multirow{5}{*}{\makecell{GPT-5\\-mini}} 
 & Thinking Token & $\text{3.836} \times 10^{-2}$ & 0.295 & \LEFTcircle  \\ 
 & Compression Rate& $\text{3.734} \times 10^{-2}$  & 0.296 &\LEFTcircle  \\  
 & Tr-Consistency (Char) &  $\text{6.668} \times 10^{-1}$  & 0.061 & \Circle \\  
 & Tr-Consistency (Token)  &  $\text{3.702} \times 10^{-3}$  & 0.415& \CIRCLE\\  
 & LLM-based judgement &  $\text{7.193} \times 10^{-4}$  & 0.488 &  \CIRCLE\\  
\specialrule{1.25pt}{0pt}{0pt}
\end{tabular}%
}
\label{tab:preliminary}
\vspace{-1.0em}
\end{table}
 
In this section, we conduct a preliminary investigation to answer the question:  \textit{Do reasoning traces contain membership signals?} 

 \vspace{0.2em}
\noindent\textbf{Experimental Setup}. The target LRMs are Claude-sonnet-4 and GPT-5-mini~\cite{openai2025gpt5}. We evaluate 200 sequences to infer membership status: 100 member sequences from books verified to appear in the training corpora of OpenAI and Claude language model series\cite{duarte2024decop,cooper2025extractingmemorizedpiecescopyrighted}, and 100 non-member sequences extracted from books published after May 2025, which are unseen by both target LRMs.

 \vspace{0.2em}
\noindent\textbf{Naive Trace-based Attacks}. We prompt the target LRMs with \textit{``What is the next word?''} of the query sequence and collect the reasoning traces. Based on the surface-level cues from the traces, we design several naive attacks:  \textbf{(1) Count of thinking tokens}: reasoning traces containing more thinking tokens (e.g., “Hmm,” “Wait,” “so”) often indicate unfamiliar, non-member sequences.
\textbf{(2) Compression rate}: The compression rate of a reasoning trace reflects how much information it contains. For unfamiliar sequences, the traces tend to involve parallel options, hesitation, and exploratory reasoning, which makes them richer in information and thus more difficult to compress. Following~\cite{morris2025languagemodelsmemorize,grunwald2004shannoninformationkolmogorovcomplexity}, it can be approximated by the per-token Negative Log-Likelihood (NLL) under GPT-2~\cite{radford2019language}. \textbf{(3) Trace consistency}: inspired by~\cite{levenshtein1966binary,dong-etal-2024-generalization}, we assume that the consistency across reasoning traces from repeated queries indicates model familiarity, which can be measured by average character-level or token-level edit distance. \textbf{(4) LLM-based judgement}: a third-party LLM (e.g., GPT-3.5-turbo~\cite{openai_gpt3_5_turbo}) is leveraged to assign an uncertainty score to the reasoning trace; higher certainty indicates members. The above metrics can serve as membership scores; an attacker predicts membership via thresholding. (See more details in Appendix~\ref{sec: naive})



\begin{figure}[t]
  \centering
  \begin{subfigure}{\columnwidth}
    \centering
    \includegraphics[width=\linewidth]{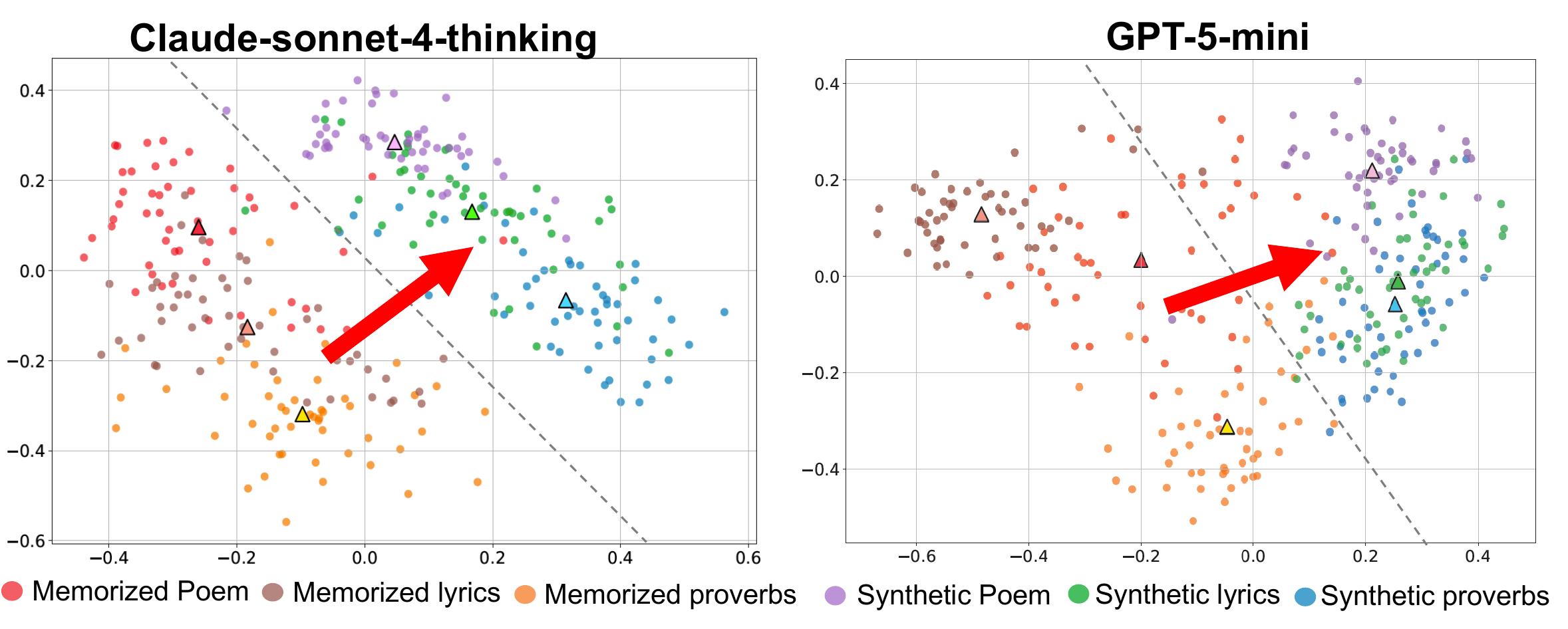}
    \caption{Extreme sample visualization across different target models}
    \label{fig:pre-claude}
    \Description{visualization}
  \end{subfigure}


  \begin{subfigure}{0.9\columnwidth}
    \centering
    \includegraphics[width=\linewidth]{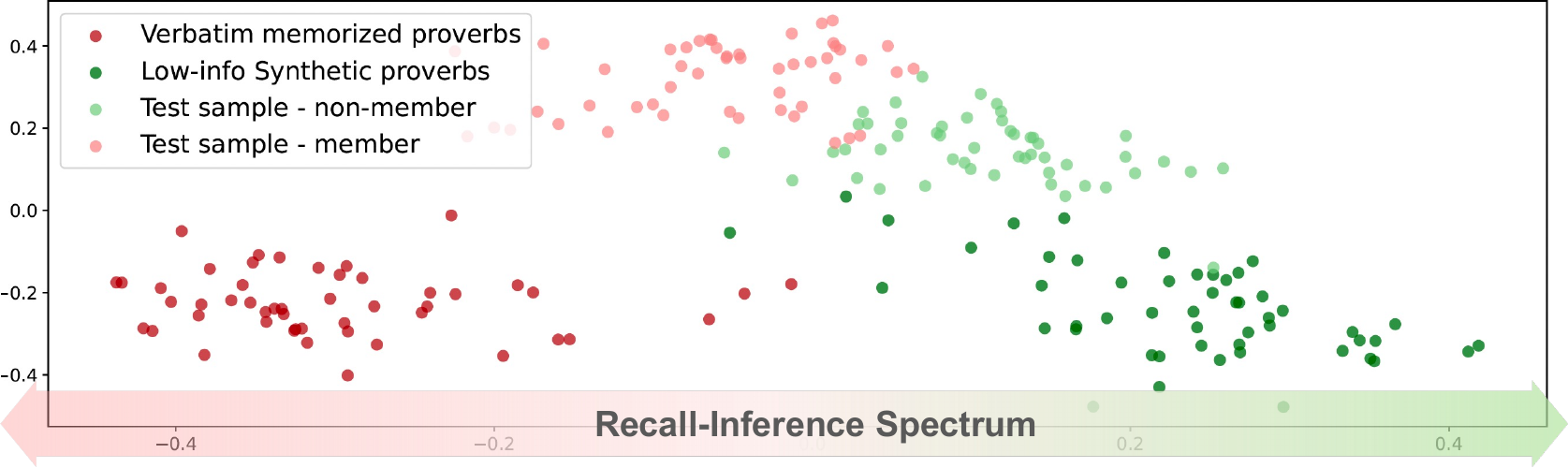}
    \caption{Recall-Inference Spectrum Phenomenon.}
    \label{fig:pre-spectrum}
  \end{subfigure}

  \caption{PCA visualization of reasoning traces.}
  \label{fig:recal-inference}
  \vspace{-1.5em}
\end{figure}

To statistically test whether the obtained score distributions of members and non-members differ significantly, we conduct a hypothesis test. Formally, let $\mu_{\text{member}}$ and $\mu_{\text{non-member}}$ denote the mean membership scores corresponding to member and non-member sequences, respectively. A two-sample, two-tailed t-test is employed to evaluate the null hypothesis~\cite{montgomery2014applied}. Formally,  we let $H_0: \mu_{\text{member}} = \mu_{\text{non-member}}$ as the null hypothesis;  $H_1: \mu_{\text{member}} \ne \mu_{\text{non-member}}$ as the alternative hypothesis. As shown in Table~\ref{tab:preliminary}, there is statistical evidence (most $p$-values are below the 0.05 significance level), providing statistical evidence to reject the null hypothesis $H_0$. Thus, we conclude that reasoning traces contain distinguishable membership signals. However, these attacks are unstable, and some scores deviate from our initial naive assumptions (e.g., on Claude-sonnet-4, trace consistency shows an opposite effect of -0.026). This instability arises because surface-level cues are highly sensitive to context and model-specific reasoning styles.


\begin{figure*}[t]
  \centering  
  \includegraphics[width=0.98\textwidth]{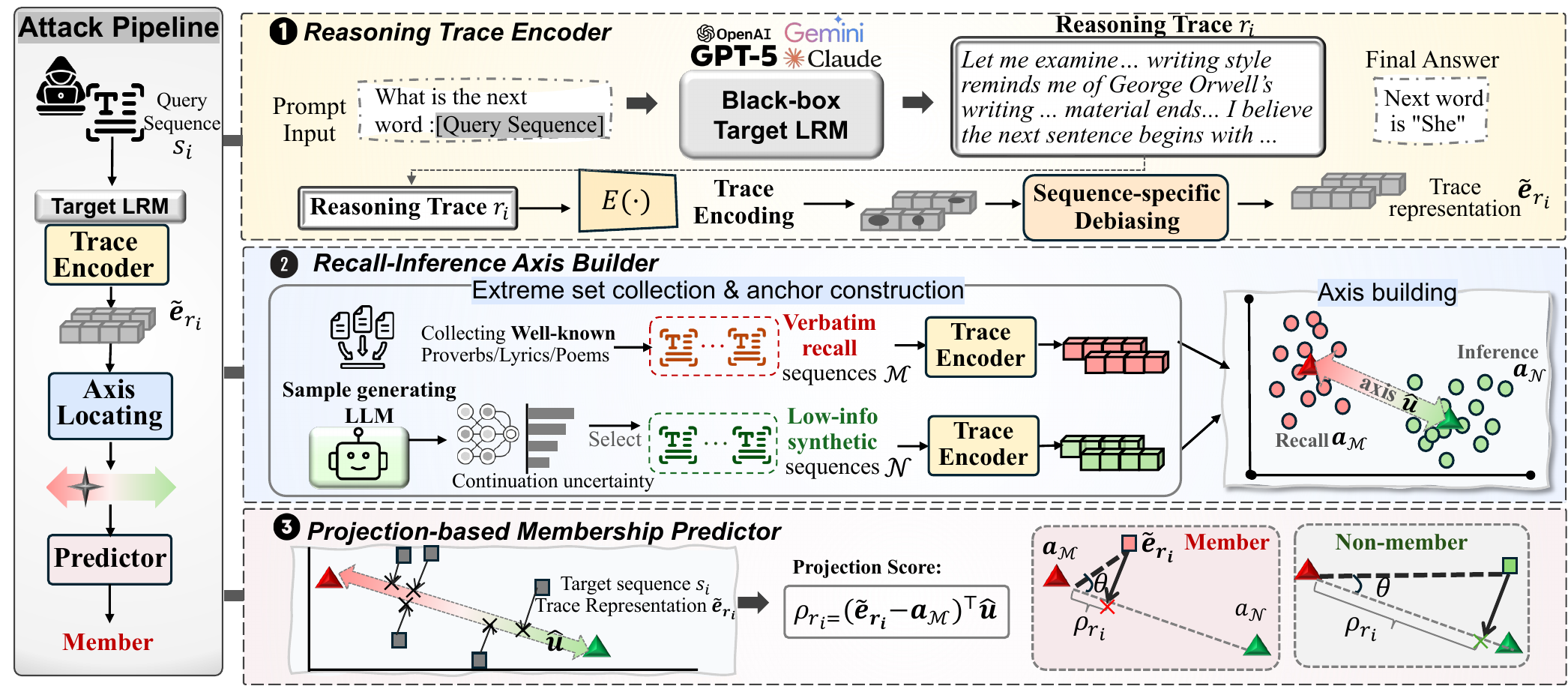}
  \caption{Overview of \textbf{BlackSpectrum}. 
 It consists of three modules: (1) Reasoning trace encoder; (2) Recall-Inference axis builder; (3) Projection-based Membership Predictor.
 }
     \Description{method}
  \label{fig: method}
    \vspace{-0.5em} 
\end{figure*}


 \vspace{0.3em}
\noindent\textbf{Trace Latent Space Observation}. To move beyond surface-level cues, we encode reasoning traces into a semantic latent space using a pretrained encoder from the sentence-transformers library~\cite{reimers-gurevych-2019-sentence}, aiming to capture deeper and stable membership patterns. We start by examining the latent-space behavior of two extreme cases: (1) highly familiar sequences that the model can reproduce verbatim, and (2) unfamiliar, low-information synthetic sequences that require the LRM to rely purely on its generalization ability to infer the continuation.

 Specifically, we construct three types of paired sequences, each containing 50 memorized sequences that most LRMs can reproduce verbatim (\emph{e.g.}, well-known proverbs, poems, and lyrics) and 50 low-information synthetic counterparts generated by GPT-3.5-turbo~\cite{openai_gpt3_5_turbo} (see Appendix Figure~\ref{fig: prompt-generate} for generation prompts). For example, the well-known proverb \textit{``An apple a day keeps the doctor…''} can be accurately reproduced by most LRMs when used as a prefix input. In contrast, a low-information synthetic prefix such as \textit{``As is often said…''} tends to make the LRM uncertain or confused.
We then encode these two extreme sequences' corresponding reasoning traces into a latent space and visualize their trace representation via Principal Component Analysis (PCA)~\cite{pearson1901pca} to observe the patterns. Figure~\ref{fig:pre-claude} shows that across different target models, the reasoning traces' representations of extreme memorized (highly familiar) and synthetic sequences (unfamiliar) are clearly separated in the latent space, forming a discernible direction. Furthermore, when we incorporate the reasoning traces of 200 book sequences into the visualization (see Figure~\ref{fig:pre-spectrum}), their trace representations lie between the two extreme groups while still exhibiting moderate separability.

In this work, we refer to the reasoning path corresponding to sequences that the LRM can reproduce verbatim as \textbf{recall-like reasoning} mode. In contrast, the reasoning path corresponding to sequences that the LRM requires must rely on generalization to predict the next word, which is referred to as \textbf{inference-like reasoning} mode. As shown in Fig.~\ref {fig:pre-spectrum}, as familiarity decreases, the trace representations gradually shift from recall-like to inference-like reasoning modes, revealing a spectrum in the latent space. We term this observation the \textbf{Recall–Inference Spectrum} phenomenon. 

%
%
%
%

\vspace{-1em}
\section{Method}\label{sec: method}

\textbf{BlackSpectrum} comprises three main modules. (1) The reasoning trace encoder extracts the reasoning traces $r_i$ corresponding to the query sequence $s_i$ and embeds them into a vector space $\tilde{\mathbf{e}}_{r_i}$. (2) The recall-inference axis builder constructs an axis  $\mathbf{\hat{u}}$ in the latent space between recall-like and inference-like reasoning modes. (3) The projection-based membership predictor obtains a membership score and predicts how likely it is to be a member.

\vspace{-0.2em}
\subsection{Reasoning Trace Encoder}

The reasoning trace encoder aims to produce the reasoning trace representation $\mathbf{e}_{r_i}$ corresponding to the query sequence $s_i$.

\begin{figure}[t]
  \centering  
  \includegraphics[width=0.98\columnwidth]{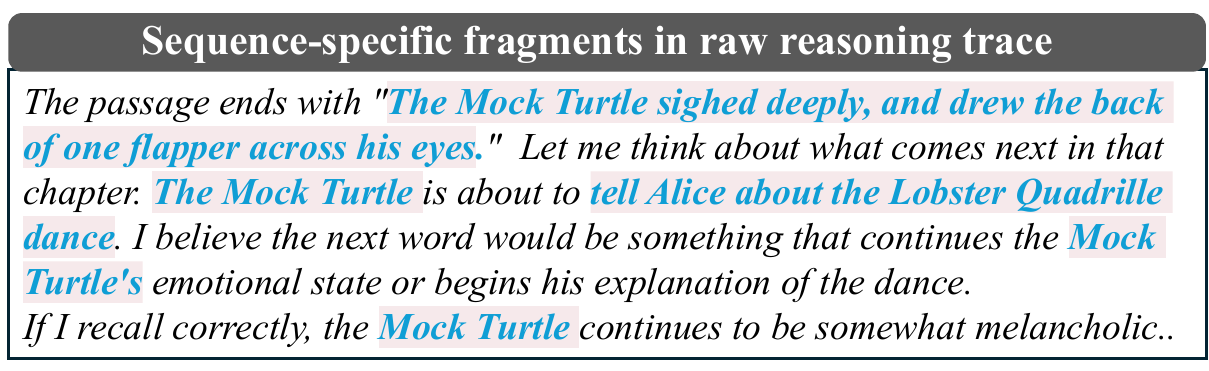}
  \caption{The case of Claude-sonnet raw reasoning trace. The unpurified reasoning trace includes fragments (highlighted) from the specific input sequence. }
  \label{fig: denoise}
    \Description{case}
   \vspace{-0.5em}
\end{figure}

 \vspace{0.3em}
\noindent\textbf{Reasoning Trace Extraction}. \label{sec: encoder}
For each sequence \( s_i\), we feed a prompt with an instruction concatenated with $s_i$ into the target LRM. And the LRM's response provides the reasoning trace $r_i$. Specifically, to ensure that the reasoning traces generated by the target model carry strong membership-related signals, the instruction should be designed to trigger behaviors that differ between members and non-members.
It should satisfy the following two conditions:
First, it explicitly encourages the target LRM to recall whether the input has been seen before.
Second, it can include a simple continuation task (\emph{e.g.,} predicting the next word), which guides the target LRM to either retrieve memorized content or generate a plausible continuation. An example of the prompt template is provided in Appendix~\ref{app: methoddetails}, Figure~\ref{fig: prompt}. 
For each sequence $s_i$, the corresponding reasoning trace $r_i$ is obtained via the target model's API.  Each reasoning trace $r_i$ is mapped by a pretrained encoder $E(\cdot)$ into a $d$-dimensional latent vector $\mathbf{e}_{r_i}$:
\vspace{-0.2em}
\begin{equation}
\begin{aligned}
\mathbf{e}_{r_i} &= E(r_i), \quad \mathbf{e}_{r_i} \in \mathbb{R}^d,
\end{aligned} \tag{2}
\end{equation}
where  $\mathbf{e}_{r_i}$ denotes the raw reasoning trace representation corresponding to the sequence $s_i$.


 \vspace{0.3em}
\noindent\textbf{Sequence-specific Denoising}. However, reasoning traces always carry the query sequence-specific signals, such as fragments of the input sequence, keywords, or stylistic patterns (see detailed example in Figure~\ref{fig: denoise}). These signals come from the target sequence itself and may interfere with the reasoning behaviors we aim to capture. To address this issue, inspired by prior work on representation debiasing~\cite{bolukbasi2016mancomputerprogrammerwoman}, we introduce a target sequence-specific debiasing module at this stage. 
Intuitively, the component of the reasoning trace representation that aligns with the input sequence representation should be removed. Technically, to separate reasoning behaviors from sequence-specific noisy content, for each reasoning trace representation $\mathbf{e}_{r_i}$ and its corresponding sequence representation $\mathbf{e}_{s_i}$, we subtract the projection of $\mathbf{e}_{r_i}$ onto $\mathbf{e}_{s_i}$:
\vspace{-0.2em}
\begin{equation}
\tilde{\mathbf{e}}_{r_i} = \mathbf{e}_{r_i} - \mathbf{p}_{r_i},
\quad \text{where} \quad
\mathbf{p}_{r_i} = \frac{\langle \mathbf{e}_{r_i}, \mathbf{e}_{s_i} \rangle}{\|\mathbf{e}_{s_i}\|^2}\,\mathbf{e}_{s_i}. \tag{3}
\end{equation}

where $\mathbf{p}_{r_i}$ denotes the sequence-specific component of $\mathbf{e}_{r_i}$. This step filters out sequence-specific interference, yielding cleaner reasoning trace embeddings that better reflect reasoning behaviors.




\subsection{Recall-Inference Axis Builder}

%
This module constructs two extreme anchors and a recall–inference axis $\mathbf{\hat{u}}$ for further measuring the target LRM’s familiarity with the query sequence $s_i$.

 \vspace{0.2em}
\noindent\textbf{Anchor Construction}. It aims to construct two groups of extreme sequences with contrasting behaviors. One is a verbatim recall set $\mathcal{M}=\{m_1,\dots,m_k\}$, consisting of sequences that most language models can reproduce exactly. The other is a low-information synthetic set $\mathcal{N}=\{n_1,\dots,n_l\}$, for which LRMs exhibit substantial uncertainty and hesitation when predicting the next token. Here, $k$ and $l$ denote the numbers of sequences in $\mathcal{M}$ and $\mathcal{N}$, respectively.
The anchors are then defined as the two centers of the reasoning trace representations of the two sets: recall-like anchor $\mathbf{a}_{\mathcal{M}}$ and inference-like anchor $\mathbf{a}_{\mathcal{N}}$. 

In particular, the sequences in the recall set $\mathcal{M}$ are drawn from well-known proverbs, song lyrics, or poems that most language models can reproduce verbatim. Yet, constructing the desired low-information sequences in the low-information synthetic set $\mathcal{N}$ is non-trivial: our desired low-information seqxuence should contain minimal guiding information within the sequence itself, thereby making the next-token continuation highly uncertain and consequently encouraging the target LRM to exhibit traces of inference-like reasoning. However, ensuring genuine uncertainty in next-token continuation remains challenging, as common linguistic patterns can still dominate the generation process.  For example, in the sequence \textit{``He poured the coffee into the \ldots''}, the word \textit{`coffee'} strongly constrains the continuation, making \textit{`cup'} a highly probable next token. In such cases, the continuation is easily predictable instead of the uncertain inference we intend to capture.

\begin{figure}[t]
  \centering  
  \includegraphics[width=0.9\columnwidth]{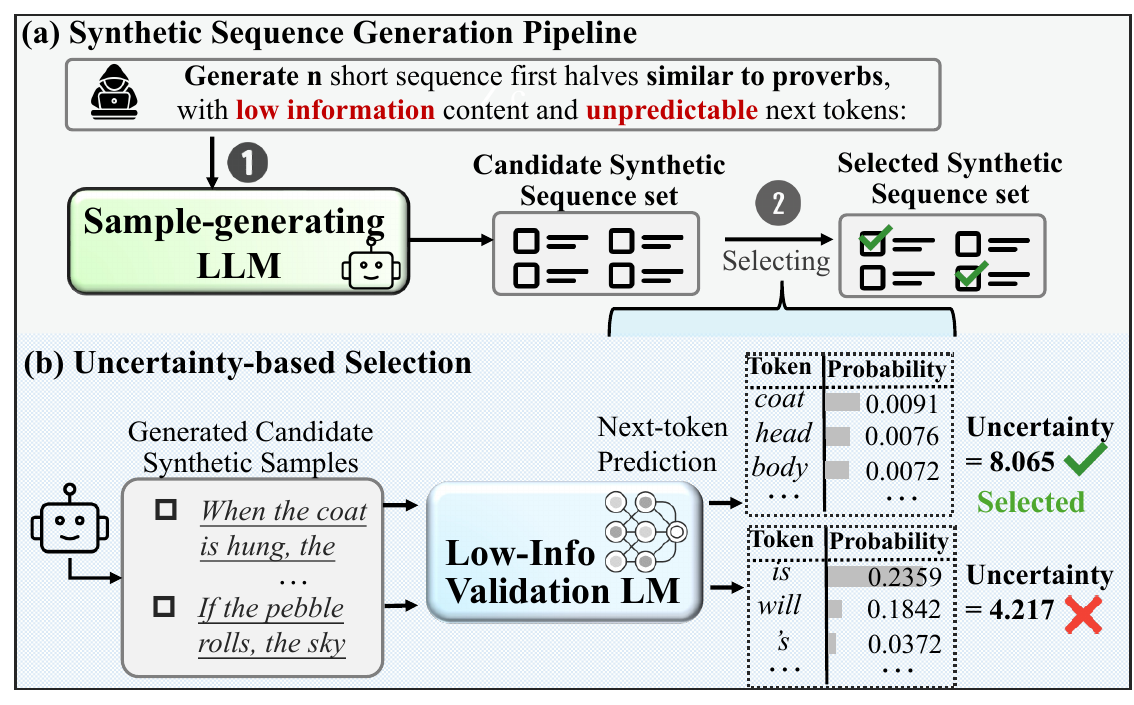}
  \caption{Synthetic sequence construction workflow.
 }
  \label{fig: synthetic}
      \Description{workflow}
   \vspace{-0.5em}
\end{figure}

\vspace{0.2em}
\noindent\textbf{Uncertainty-based Synthetic Sequence Selection}. Therefore, for low-information synthetic sequence construction, we design a two-step workflow,  as illustrated in Figure~\ref{fig: synthetic}. First, a sample-generating LLM is introduced to generate a small set of candidate synthetic sequences. Second, we design an uncertainty-based selection algorithm to filter out unexpected low-information synthetic sequences. Specifically, as shown in Figure~\ref{fig: synthetic} (b), a low-information validation language model (validation LM) is employed to perform next-token prediction based on the candidate synthetic sequence $n_i$. Following previous work~\cite{shannon}, we quantify the continuation uncertainty by computing the Shannon entropy of the next-token distribution: 

\begin{equation}
H(n_i) = - \sum_{w \in V} P(w \mid n_i) \log P(w \mid n_i), \tag{4}
\end{equation}

where $V$ denotes the vocabulary set of the validation LM, and $P(w \mid n_i)$ is the probability it assigns to token $w$ given prefix $n_i$. Top-$\gamma$ highest-entropy sequences are selected as final synthetic sequences, as they indicate the most uncertain continuations that require full inference. From the perspective of LRMs, high-entropy continuation results in reasoning traces characterized by diverse inference trajectories and divergent thinking.

%

%

Based on the above obtained sequences,  the anchors are created from the centers of their reasoning-trace representations. Specifically, we obtain these representations $\mathbf{\tilde{e}}_{m_i}$ and $\mathbf{\tilde{e}}_{n_i}$ using the reasoning trace encoder introduced in Section~\ref{sec: encoder}, which consists of encoding and denoising steps. Given the verbatim recall sequence set $\mathcal{M}$ and the low-information synthetic sequence set $\mathcal{N}$, the  recall-like anchor and inference-like anchor can be computed as:  
\begin{equation}
\mathbf{a}_{\mathcal{M}} = \frac{1}{k} \sum_{i=1}^{k} \mathbf{\tilde{e}}_{m_i}, 
\quad 
\mathbf{a}_{\mathcal{N}} = \frac{1}{l} \sum_{i=1}^{l} \mathbf{\tilde{e}}_{n_i},\tag{5}
\end{equation}
where $ \mathbf{\tilde{e}}_{m_i}$ and $ \mathbf{\tilde{e}}_{n_i}$ denote the denoised reasoning-trace representations obtained from the trace encoder, and $k$ and $l$ indicate the number of sequences in the sets, respectively.
%

 \vspace{0.2em}
\noindent\textbf{Axis Building}. Following the derivation of the recall-like and inference-like anchor vectors, an axis is built to capture the contrast between the two distinct reasoning modes. Specifically, the axis direction is defined as the vector from the recall-like anchor $a_{\mathcal{M}}$ to the inference-like anchor $a_{\mathcal{N}}$. It can be formalized as follows: 
\begin{equation}
\mathbf{u} = \mathbf{a}_{\mathcal{N}} - \mathbf{a}_{\mathcal{M}}, \quad 
\mathbf{\hat{u}} = \frac{\mathbf{u}}{\|\mathbf{u}\|}, \tag{6}
\end{equation}
where $\mathbf{\hat{u}}$ is the normalized axis direction, providing a reference direction for contrasting recall and inference reasoning modes.  

\subsection{Projection-based Membership Predictor}
With the recall-like anchor $a_{\mathcal{M}}$, the inference-like anchor $a_{\mathcal{N}}$, and the axis $\hat{\mathbf{u}}$ connecting them established, a projection-based predictor is then designed to predict the membership status $m_i$ of the query sequence $s_i$.

 \vspace{0.2em}
\noindent\textbf{Projection Score}. 
Given a target sequence $s_i$, we first obtain its denoised reasoning trace embedding $\tilde{e}_{r_i}$ 
using the trace encoder. 
The embedding is then projected onto the recall--inference axis $\hat{\mathbf{u}}$, 
which can be formalized as:
\begin{equation}
\rho_{r_i} = (\mathbf{\tilde{e}}_{r_i} - \mathbf{a}_{\mathcal{M}})^{\top} \hat{\mathbf{u}}, \tag{7}
\end{equation}
where $\rho_{r_i}$ is the scalar projection of the denoised reasoning trace embedding 
$\mathbf{\tilde{e}}_{r_i}$ of the query sequence $s_i$ onto the recall--inference axis $\hat{\mathbf{u}}$. Sequences whose projections fall closer to the recall anchor $\mathbf{a}_{\mathcal{M}}$ are more likely to behave as members. (See Figure~\ref{fig: method})

 \vspace{0.2em}
\noindent\textbf{Membership Predictor}.  Following the previous MIA paradigm~\cite{zhang2025mink,duarte2024decop}, we first define a membership score for each target sequence and compare it to a pre-defined threshold for distinguishing members and non-members. Technically, to define the membership score $\epsilon_{s_i}$, we normalize the projection of the sequence's ($s_i$) reasoning trace $\rho_{r_i}$  by the distance between the two anchors. The distance can be denoted as $D = \|\textbf{a}_{\mathcal{N}} - \textbf{a}_{\mathcal{M}}\|$. This normalization establishes an intuitive reference scale along the recall–inference axis, where the recall-like anchor is mapped to $1$ and the inference-like anchor to $0$. Formally, the membership score is defined as:
\begin{equation}
\epsilon_{s_i} = 1 - \frac{\rho_{r_i}}{D}.\tag{8}
\end{equation}
Higher values indicate recall-like reasoning modes (member-like), while lower values indicate inference-like reasoning modes (non-member-like).


\section{Experiments}
\begin{table}[t]
   \caption{Statistics of the \textit{arXivReasoning} and \textit{BookReasoning}.}
   \centering
   \Large
   \renewcommand{\arraystretch}{0.7}
   \resizebox{0.92\columnwidth}{!}{
   \begin{tabular}{lcccc}
       \toprule
       \textbf{Dataset} & \textbf{Type} & \textbf{Seq. Length} & \textbf{\#Sequences} & \textbf{\#Total} \\
       \midrule
       \multirow{2}{*}{arXivReasoning} 
         & Member     & 32 / 64 / 128 & 761  & \multirow{2}{*}{1605} \\ 
         & Non-member & 32 / 64 / 128 & 844  &  \\ 
       \midrule
       \multirow{2}{*}{BookReasoning} 
         & Member     & 32 / 64 / 128 & 3505 & \multirow{2}{*}{6112} \\
         & Non-member & 32 / 64 / 128 & 2607 &  \\
       \bottomrule
   \end{tabular}}
   \label{tab:dataset}
   \vspace{-0.5em}
\end{table}

\subsection{Experimental Settings}
\begin{table*}[ht]
\centering
\renewcommand{\arraystretch}{0.9}  
\vspace{\baselineskip}
\caption{Attack Performance.  \#Q indicates how many API queries are required for each sample. For each target model, best results are bold and \underline{underlined}; second-best are underlined.  $\Delta$Attack Performance (\%) measures relative gains over the best prior attack that \emph{do not} use reasoning traces.
(\textcolor{red}{$\uparrow$}: improvement; \textcolor{green}{$\downarrow$}: fewer queries).}
\label{tab:sentenceperformance}
\Huge
\resizebox{\textwidth}{!}{%
\begin{tabular}{@{}c|c|ccc|ccc|ccc|ccc|ccc@{}}
\Xhline{1.5pt}  

\multicolumn{17}{c}{{Sequence-level Results}} \\
\Xhline{1pt}

 \multirow{2}{*}{\textbf{arXivReasoning}} & \multirow{2}{*}{\textbf{\makecell{\# Q} $\downarrow$}} & 
 \multicolumn{3}{c}{\textbf{QwQ-32B}} & 
 \multicolumn{3}{c}{\textbf{Gemini-2.5-flash}} & 
 \multicolumn{3}{c}{\textbf{Claude-sonnet-4}} & 
 \multicolumn{3}{c}{\textbf{GPT-5-mini}} &
 \multicolumn{3}{c}{\textbf{Avg Performance}} \\ 
 \cmidrule(lr){3-5}\cmidrule(lr){6-8}\cmidrule(lr){9-11}\cmidrule(lr){12-14}\cmidrule(lr){15-17}
 && ACC & AUC & T@5\%F & ACC & AUC & T@5\%F & ACC & AUC & T@5\%F & ACC & AUC & T@5\%F & ACC & AUC & T@5\%F \\ 
\midrule
SaMIA~\cite{kaneko2024samplingbasedpseudolikelihoodmembershipinference} & 5 & 
0.549 &0.551& 0.056&0.500 &0.405& 0.005 &  0.523  & 0.521 & 0.060& 0.523  & 0.511 &0.033 &0.524 &0.497&0.039
 \\
CDD~\cite{dong-etal-2024-generalization} & 5 & 
0.500  &  0.451  &  0.040  & 0.500 &  0.376  & 0.026  &\underline{0.553} &\underline{0.555}  & \underline{0.114}&0.513& 0.503  & \underline{0.054} & 0.516 &0.471 & 0.059 \\
DE-COP~\cite{duarte2024decop} & 24 & 
\underline{0.620} &  \underline{0.650} &  0.080  &0.509  & 0.503  &  0.050 &  --  & ---  & --&---  & ---  & --- & 0.565 & 0.577 & 0.065 \\
\cmidrule[0.1pt](lr){1-17}
Thinking Token \textbf{\textit{(ours)}} & 1 & 
0.505 &  0.420 &  0.058  &0.549  &  0.564& 0.060& 0.506 & 0.492  & 0.041 &0.503  & 0.448 & 0.040 & 0.516& 0.481 & 0.050 \\
Compression Rate \textbf{\textit{(ours)}} & 1 & 
0.503 & 0.397 &  0.037 & 0.568&  0.583  & 0.093  & 0.534 & 0.540  & 0.087&0.505  & 0.399  & 0.022 &0.528 &0.480 &0.060 \\
Tr-Consistency (Char) \textbf{\textit{(ours)}} & 3 & 
0.501  &  0.409 &  0.014  &0.561 & 0.571  &  0.088 &  0.525& 0.522&0.051 & 0.541 &0.541  & 0.042 & 0.532& 0.511& 0.049\\ 
Tr-Consistency (Token) \textbf{\textit{(ours)}} & 3 & 
0.510 &  0.465 &  0.018 &0.573 & 0.582  &  0.066 & 0.545& 0.544 &  0.073& \underline{0.553} & \underline{0.556}  & 0.031 & 0.545&0.537 & 0.047\\
LLM-based judgement  \textbf{\textit{(ours)}}& 2 & 
0.600 & 0.584  & \underline{ 0.139}  & \underline{0.596} & \underline{0.537} & \underline{0.114}  & 0.518& 0.379 & 0.074 &0.517 & 0.494  & 0.047 & 0.558 &0.499 &0.094 \\
\textbf{BlackSpectrum} \textbf{\textit{(ours)}}& \underline{\bf{1}}& 
\underline{\bf{0.680}}  &\underline{\bf{0.736}}  & \underline{\bf{0.278}} & \underline{\bf{0.664}}&\underline{\bf{0.721}}& \underline{\bf{0.188}} &\underline{\bf{0.671}} & \underline{\bf{0.733}} & \underline{\bf{0.189}} &\underline{\bf{0.565}}  & \underline{\bf{0.585}}  & \underline{\bf{0.076}} & 
\underline{\bf{0.645}} & \underline{\bf{0.694}} & \underline{\bf{0.183}} \\
\cmidrule[0.1pt](lr){1-17}
 \rowcolor{gray!20} 
\textbf{$\Delta$ Attack Performance} \% & $\approx$1/5\textcolor{green}{$\downarrow$}& 9.7\%\textcolor{red}{$\uparrow$} & 13.2\%\textcolor{red}{$\uparrow$} & 2.5$\times$\textcolor{red}{$\uparrow$}  &30.5\%\textcolor{red}{$\uparrow$} &  43.3\%\textcolor{red}{$\uparrow$}&   2.8$\times$\textcolor{red}{$\uparrow$} &  21.3\%\textcolor{red}{$\uparrow$}&  32.1\%\textcolor{red}{$\uparrow$} & 65.8\%\textcolor{red}{$\uparrow$}& 8.0\%\textcolor{red}{$\uparrow$}  & 14.5\%\textcolor{red}{$\uparrow$} & 40.7\%\textcolor{red}{$\uparrow$}  &14.2\%\textcolor{red}{$\uparrow$}  & 20.2\%\textcolor{red}{$\uparrow$} & 1.8$\times$\textcolor{red}{$\uparrow$}\\
\hline\hline

 \multirow{2}{*}{\textbf{BookReasoning}} & \multirow{2}{*}{\textbf{\makecell{\# Q} $\downarrow$}} & 
 \multicolumn{3}{c}{\textbf{QwQ-32B}} & 
 \multicolumn{3}{c}{\textbf{Gemini-2.5-flash}} & 
 \multicolumn{3}{c}{\textbf{Claude-sonnet-4}} & 
 \multicolumn{3}{c}{\textbf{GPT-5-mini}} &
 \multicolumn{3}{c}{\textbf{Avg Performance}} \\ 
 \cmidrule(lr){3-5}\cmidrule(lr){6-8}\cmidrule(lr){9-11}\cmidrule(lr){12-14}\cmidrule(lr){15-17}
&& ACC & AUC & T@5\%F & ACC & AUC & T@5\%F & ACC & AUC & T@5\%F & ACC & AUC & T@5\%F & ACC & AUC & T@5\%F \\ \midrule
SaMIA~\cite{kaneko2024samplingbasedpseudolikelihoodmembershipinference} & 5 &
0.549 &0.544& 0.067  &  0.522 & 0.515  & 0.065 & 0.531& 0.529 & 0.107&0.559& 0.556 & \underline{0.103} &0.540 & 0.536& 0.086 \\
CDD~\cite{dong-etal-2024-generalization} & 5 &
0.501& 0.442 & 0.049  & 0.568&  0.567  &  0.032  & 0.548 & 0.526  &0.070&0.548&0.526  &0.070 & 0.541 & 0.515 & 0.055 \\
\cmidrule[0.1pt](lr){1-17}
Thinking Token \textbf{\textit{(ours)}} & 1 & 
0.573  &  0.574  &  0.092  & 0.573 &  0.592&  0.070 & 0.576 & 0.591 &0.103 &0.505  & 0.476  & 0.047 & 0.557 & 0.558& 0.078 \\
Compression Rate  \textbf{\textit{(ours)}} & 1 & 
0.570  &  0.572 & 0.071 & 0.576 &  0.595&  0.084  & 0.578 & 0.601  & 0.074 &0.566  & 0.575  & 0.045 &0.573 & 0.586 & 0.069 \\
Tr-Consistency (Char)  \textbf{\textit{(ours)}} & 3 & 
0.583  &  0.548 & \underline{0.197} & 0.533& 0.540 &   0.080  &0.541 & 0.546 & 0.059&0.517 & 0.489  & 0.047 &0.544 & 0.531 & 0.096 \\ 
Tr-Consistency (Token)  \textbf{\textit{(ours)}} & 3& 
0.594 & 0.577&\underline{\bf{0.200}} &  0.502 & 0.489  &0.035  &0.595 &\underline{0.622} & 0.103&0.517 & 0.490  & 0.059 & 0.552& 0.545& 0.099 \\
LLM-based judgement  \textbf{\textit{(ours)}}& 2 & 
0.549&0.524 & 0.098 & \underline{0.603} &\underline{0.665} &  \underline{0.233}  &\underline{0.596}& 0.579& \underline{0.115} &\underline{0.611}  &\underline{0.634} & 0.087 & 0.590& 0.601 & 0.133 \\
\textbf{BlackSpectrum}  \textbf{\textit{(ours)}}&\underline{\textbf{1}} & 
\underline{\bf{0.781}}  &  \underline{\bf{0.735}}  &  0.192  &\underline{\bf{0.698}}& \underline{\bf{0.754}}  & \underline{\bf{0.249}} & \underline{\bf{0.727}}& \underline{\bf{0.795}}& \underline{\bf{0.287}}&\underline{\bf{0.677}}& \underline{\bf{0.707}} & \underline{\bf{0.270}} &
\underline{\bf{0.721}} & \underline{\bf{0.748}} & \underline{\bf{0.250}} \\
\cmidrule[0.1pt](lr){1-17}
 \rowcolor{gray!20} 
\textbf{$\Delta$ Attack Performance}\% & $\approx$1/5\textcolor{green}{$\downarrow$}&42.2\%\textcolor{red}{$\uparrow$}&  35.1\%\textcolor{red}{$\uparrow$} & 1.9$\times$\textcolor{red}{$\uparrow$}& 22.9\%\textcolor{red}{$\uparrow$} &   32.9\%\textcolor{red}{$\uparrow$}  & 2.8$\times$\textcolor{red}{$\uparrow$} & 32.7\%\textcolor{red}{$\uparrow$}  & 50.3\%\textcolor{red}{$\uparrow$} &1.7$\times$\textcolor{red}{$\uparrow$}& 21.1\%\textcolor{red}{$\uparrow$}  & 27.2\%\textcolor{red}{$\uparrow$}  &1.6$\times$\textcolor{red}{$\uparrow$}&33.3\%\textcolor{red}{$\uparrow$}  & 39.6\%\textcolor{red}{$\uparrow$} & 1.9$\times$\textcolor{red}{$\uparrow$} \\

\Xhline{1.2pt}
\multicolumn{17}{c}{Document-level Results} \\
\Xhline{1.2pt}

\multirow{2}{*}{\textbf{arXivReasoning}} &
\multirow{2}{*}{\textbf{\makecell{\# Q} $\downarrow$}} &
\multicolumn{3}{c}{\textbf{QwQ-32B}} &
\multicolumn{3}{c}{\textbf{Gemini-2.5-flash}} &
\multicolumn{3}{c}{\textbf{Claude-sonnet-4}} &
\multicolumn{3}{c}{\textbf{GPT-5-mini}} &
\multicolumn{3}{c}{\textbf{Avg Performance}} \\ 

\cmidrule(lr){3-5}\cmidrule(lr){6-8}\cmidrule(lr){9-11}\cmidrule(lr){12-14}\cmidrule(lr){15-17}
& & ACC & AUC & T@5\%F & ACC & AUC & T@5\%F & ACC & AUC & T@5\%F & ACC & AUC & T@5\%F & ACC & AUC & T@5\%F \\ 

\midrule
SaMIA~\cite{kaneko2024samplingbasedpseudolikelihoodmembershipinference}& 5 &
0.657 & 0.656 & 0.120 & 0.500 & 0.039 & 0.000 & 0.610 & 0.620 & \underline{0.200} & 0.573 & 0.492 & 0.000 &0.585 & 0.452 & 0.080 \\

CDD~\cite{dong-etal-2024-generalization}& 5 &
0.500 & 0.295 & 0.000 & 0.643 & 0.594 & 0.000 & \underline{0.820} & \underline{0.800} & 0.160 & 0.580 & 0.517 & \underline{0.160} & 0.636& 0.662 & 0.080 \\

DE-COP~\cite{duarte2024decop}&24&
\underline{0.840} & \underline{0.926} & \underline{0.640} & 0.567 & 0.456 & 0.060 & -- & -- & -- & -- & -- & -- & 0.704 & 0.691&0.350 \\

\cmidrule[0.1pt](lr){1-17}
Thinking Token \textbf{\textit{(ours)}} & 1 &
0.523 & 0.240 & 0.080 & 0.760 & 0.805 & 0.200 & 0.530 & 0.461 & 0.000 & 0.517 & 0.291 & 0.000 & 0.583 & 0.449 & 0.070 \\

Compression Rate \textbf{\textit{(ours)}} & 1 &
0.500 & 0.139 & 0.000 & 0.700 & 0.710 & 0.200 & 0.663 & 0.665 & 0.040 & 0.500 & 0.168 & 0.000 & 0.591 &0.421& 0.060 \\

Tr-Consistency (Char) \textbf{\textit{(ours)}} & 3 &
0.500 & 0.149 & 0.000 & \underline{0.770} & \underline{0.816} & \underline{0.360} & 0.603 & 0.592 & 0.012 & 0.697 & 0.667 & \underline{\bf{0.280}} & 0.665 & 0.556 &0.163 \\

Tr-Consistency (Token) \textbf{\textit{(ours)}} & 3 &
0.533 & 0.339 & 0.000 & 0.737 & 0.747 & 0.080 & 0.676 & 0.676 & 0.040 & 0.670 & 0.652 & 0.120 & 0.654& 0.604 & 0.060 \\

LLM-judgement \textbf{\textit{(ours)}} & 2 &
0.763 & 0.716 & 0.560 & 0.580 & 0.487 & 0.080 & 0.500 & 0.077 & 0.000 & 0.550 & 0.412 & 0.000 & 0.598 & 0.423 & 0.160\\

\textbf{BlackSpectrum} \textbf{\textit{(ours)}} &\underline{\textbf{1}} &
\underline{\bf{0.947}} & \underline{\bf{0.965}} & \underline{\bf{0.922}} &
\underline{\bf{0.950}} & \underline{\bf{0.977}} & \underline{\bf{0.880}} &
\underline{\bf{0.863}} & \underline{\bf{0.901}} & \underline{\bf{0.560}} &
\underline{\bf{0.724}} & \underline{\bf{0.731}} & 0.115 &
\underline{\bf{0.871}} & \underline{\bf{0.894}} & \underline{\bf{0.619}} \\

\cmidrule[0.1pt](lr){1-17}
\rowcolor{gray!20} 
\textbf{$\Delta$ Attack Performance} \% & $\approx$1/5 \textcolor{green}{$\downarrow$} & 12.7\% \textcolor{red}{$\uparrow$} & 4.2\%\textcolor{red}{$\uparrow$} &44.1\%\textcolor{red}{$\uparrow$} & 47.7\%\textcolor{red}{$\uparrow$} & 64.5\%\textcolor{red}{$\uparrow$}&13.6$\times$\textcolor{red}{$\uparrow$} &  5.2\%\textcolor{red}{$\uparrow$} & 12.6\%\textcolor{red}{$\uparrow$} &12.8$\times$\textcolor{red}{$\uparrow$} &24.8\%\textcolor{red}{$\uparrow$} & 41.4\%\textcolor{red}{$\uparrow$} & 75.0\%\textcolor{red}{$\uparrow$}& 23.7\%\textcolor{red}{$\uparrow$}& 29.4\%\textcolor{red}{$\uparrow$} & 50.7\%\textcolor{red}{$\uparrow$} \\
\hline\hline

\multirow{2}{*}{\textbf{BookReasoning}} &
\multirow{2}{*}{\textbf{\makecell{\# Q} $\downarrow$}} &
\multicolumn{3}{c}{\textbf{QwQ-32B}} &
\multicolumn{3}{c}{\textbf{Gemini-2.5-flash}} &
\multicolumn{3}{c}{\textbf{Claude-sonnet-4}} &
\multicolumn{3}{c}{\textbf{GPT-5-mini}} &
\multicolumn{3}{c}{\textbf{Avg Performance}} \\

\cmidrule(lr){3-5}\cmidrule(lr){6-8}\cmidrule(lr){9-11}\cmidrule(lr){12-14}\cmidrule(lr){15-17}
& & ACC & AUC & T@5\%F & ACC & AUC & T@5\%F & ACC & AUC & T@5\%F & ACC & AUC & T@5\%F & ACC & AUC & T@5\%F \\ 

\midrule
SaMIA~\cite{kaneko2024samplingbasedpseudolikelihoodmembershipinference}& 5 &
0.638 & 0.669 & 0.150 & 0.700 & 0.610 & 0.400 & 0.750 & 0.775 & 0.300 & \underline{0.850} & \underline{0.860} & 0.300 &0.735 & 0.729&0.288 \\

CDD~\cite{dong-etal-2024-generalization}& 5 &
0.507 & 0.286 & 0.027 & 0.600 & 0.530 & 0.000 & 0.750 & 0.690 & 0.100 & 0.800 & 0.845 & 0.400 & 0.664& 0.588 &0.132 \\

\cmidrule[0.1pt](lr){1-17}
Thinking Token \textbf{\textit{(ours)}} & 1 &
0.729 & 0.718 & \underline{0.362} & 0.751 & 0.812 & 0.486 & 0.756 & \underline{0.839} & 0.447 & 0.500 & 0.419 & 0.010 & 0.684 &0.697 &0.326 \\

Compression Rate \textbf{\textit{(ours)}} & 1 &
\underline{0.734} & 0.734 & 0.295 & 0.710 & 0.794 & 0.352 & 0.697 & 0.733 & 0.219 & 0.698 & 0.709 & 0.038 &0.710& 0.742& 0.226 \\

Tr-Consistency (Char) \textbf{\textit{(ours)}} & 3 &
0.599 & 0.592 & 0.153 & 0.686 & 0.733 & 0.304 & 0.750 & 0.670 & 0.010 & 0.500 & 0.320 & 0.000 &0.634 & 0.579 & 0.117 \\

Tr-Consistency (Token) \textbf{\textit{(ours)}} & 3 &
0.658 & \underline{0.784} & 0.305 & 0.506 & 0.400 & 0.029 & \underline{0.850} & 0.790 & \underline{0.700} & 0.500 & 0.350 & 0.000 & 0.629 & 0.581 & 0.259 \\

LLM-judgement \textbf{\textit{(ours)}} & 2 &
0.718 & 0.719 & \underline{\bf{0.390}} & \underline{0.894} & \underline{0.923} & \underline{0.760} & 0.760 & 0.790 & 0.533 & 0.816 & 0.836 & \underline{0.476} & 0.797 & 0.817&0.540 \\

\textbf{BlackSpectrum} \textbf{\textit{(ours)}} &\underline{\textbf{1}}&
\underline{\bf{0.937}} & \underline{\bf{0.902}} & 0.324 &
\underline{\bf{0.900}} & \underline{\bf{0.941}} & \underline{\bf{0.838}} &
\underline{\bf{0.946}} & \underline{\bf{0.983}} & \underline{\bf{0.867}} &
\bf{0.892} & \bf{0.912} & \bf{0.686} &
\underline{\bf{0.919}} & \underline{\bf{0.935}} & \underline{\bf{0.679}} \\

\cmidrule[0.1pt](lr){1-17}
\rowcolor{gray!20} 
\textbf{$\Delta$ Attack Performance} \% & $\approx$1/5 \textcolor{green}{$\downarrow$} & 27.7\%\textcolor{red}{$\uparrow$} & 34.8\%\textcolor{red}{$\uparrow$}& 1.6$\times$ \textcolor{red}{$\uparrow$} & 28.6\%\textcolor{red}{$\uparrow$} & 54.3\%\textcolor{red}{$\uparrow$}& 52.3\%\textcolor{red}{$\uparrow$} & 26.1\%\textcolor{red}{$\uparrow$}&26.8\%\textcolor{red}{$\uparrow$} &  1.89$\times$ \textcolor{red}{$\uparrow$} &4.9\%\textcolor{red}{$\uparrow$} & 6.0\%\textcolor{red}{$\uparrow$}& 71.5\%\textcolor{red}{$\uparrow$}  &25.0\%\textcolor{red}{$\uparrow$}  &28.3\%\textcolor{red}{$\uparrow$}& 1.4$\times$\textcolor{red}{$\uparrow$} \\
\Xhline{1.5pt} 
\end{tabular}}
\end{table*}

\begin{table}[t]
\centering
\renewcommand{\arraystretch}{0.9}
\vspace{\baselineskip}
\caption{Ablation studies.}
\vspace{-0.5em}
\large
\label{tab:ablation}
\resizebox{0.95\columnwidth}{!}{%
\begin{tabular}{l|ccc}
\specialrule{1.1pt}{0pt}{0pt}
\textbf{Module} & \textbf{ACC} & \textbf{AUC} & \textbf{T@5\%F} \\ 
\midrule
w/o denoising module  & 0.583 \textcolor{gray}{(-0.081)} & 0.608 \textcolor{gray}{(-0.113)} & 0.101 \textcolor{gray}{(-0.087)} \\
\cmidrule[0.1pt](lr){1-4}
all-MiniLM-L6-v2      & 0.664  & 0.721 & 0.188 \\
all-MiniLM-L12-v2     & 0.664  & 0.721 & 0.188 \\
all-distilroberta-v1  & 0.678  & 0.727 & 0.216 \\
\midrule
w/o selection module  & 0.659 \textcolor{gray}{(-0.005)} & 0.711 \textcolor{gray}{(-0.010)} & 0.152 \textcolor{gray}{(-0.013)} \\
\midrule
\textbf{Full}         & \textbf{0.664} & \textbf{0.721} & \textbf{0.188} \\
\specialrule{1.1pt}{0pt}{0pt}
\end{tabular}
}
\end{table}

\subsubsection{Datasets and Target Models} \label{sec: datasets}


Existing MIA datasets, originally developed for pre-training language models, are no longer suitable for modern LRMs.
Many of the materials contained in these datasets may have already been included in the training corpora of recent emerging LRMs, undermining their validity as evaluation benchmarks.  That is, when targeting modern LRMs, the member–non-member labeling is unreliable. To address the above limitations, we construct two new datasets specific to commercial black-box LRMs:  \textbf{arXivReasoning} and \textbf{BookReasoning}. The statistics are shown in Table~\ref{tab:dataset}. In particular, for arXivReasoning, it contains 1605 sequences extracted from 55 arXiv papers.  For BookReasoning, it contains 6112 sequences extracted from 184 books. For both datasets, non-members are constructed from materials published after May 2025 to ensure that they remain unseen by most LRMs. Members are drawn from the labeled member split commonly used in prior work~\cite{duarte2024decop, shi2024detecting, hu2025ijcai} and are reused as the member data in our newly constructed datasets.
(See Appendix~\ref{app: datasetdetails} for details on data collection and statistics.)  Our target models are the advanced commercial LRMs: \textbf{QwQ-32B}~\cite{qwen2025qwq32b},\textbf{Gemini-2.5-flash} ~\cite{comanici2025gemini25pushingfrontier}, \textbf{Claude-sonnet-4}~\cite{anthropic2025claude4}, and \textbf{GPT-5-mini}~\cite{openai2025gpt5}. We note that this setting may be imperfect due to limited knowledge of the exact training data of some closed-source LLMs. Yet, consistent with prior work~\cite{duarte2024decop, shi2024detecting}, this setting is widely adopted and adequate for empirical evaluation.

\vspace{-0.4em}

\subsubsection{Evaluation Metrics}
Following prior MIA studies~\cite{carlini2021extracting,carlini2022membershipinferenceattacksprinciples,chen-etal-2025-statistical}, we use three metrics:  Balanced Accuracy (ACC) measures the average attack accuracy over a balanced dataset;  Area Under the ROC Curve (AUC) captures the overall attack performance across thresholds. True Positive Rate at low False Positive Rate (TPR@5\%FPR) evaluates the attacks' sensitivity under strict false positive constraints.


\vspace{-0.4em}
\subsubsection{Baseline Methods}
Most existing MIA methods~\cite{shi2024detecting,meeus2023did,carlini2021extracting} rely on token logits and are thus inapplicable to commercial LRMs, so we adopt three black-box methods originally designed for pre-training LLMs as baselines for attacking modern LRMs.
Specifically, \textbf{SaMIA}~\cite{kaneko2024samplingbasedpseudolikelihoodmembershipinference} evaluates membership by measuring the ROUGE~\cite{lin2004rouge} overlap between multiple suffixes generated by the language model and the actual suffix of the sequence; \textbf{CDD}~\cite{dong-etal-2024-generalization} also depends on multiple continuous suffixes and measures token-level edit distance among the continuous suffixes themselves; \textbf{DE-COP}~\cite{duarte2024decop} evaluates membership of copyright books by constructing well-designed multiple-choice questions.  A target model that answers more questions correctly is regarded as more likely to have seen the book during training.
Moreover, we include five naive reasoning-trace-based baselines: \textbf{Thinking Tokens}, \textbf{Compression Rate}, \textbf{Tr-Consistency (Char)}, \textbf{Tr-Consistency (Token)}, and \textbf{LLM-based judgement}. Details are provided in Appendix~\ref{sec: naive}.

\vspace{-0.5em}
\subsubsection{Implemental Details}
All experiments are conducted on a workstation equipped with NVIDIA Tesla V100 GPUs, together with the commercial real-world APIs.
We leverage the pre-trained sentence embedding model from the sentence-transformers library~\cite{reimers-gurevych-2019-sentence} (\emph{e.g.}, all-MiniLM-L6-v2, all-distilroberta-v1, \emph{etc.}) Our main experiments construct 50 well-known proverbs. And use the sample-generating LLM (\textit{GPT-3.5-turbo}~\cite{openai_gpt3_5_turbo}) to first generate 100 proverb-like synthetic sequences, from which $\gamma=50$ low-information samples are selected for subsequent analysis by validation LM (\textit{GPT-2}~\cite{radford2019language}). All target LRMs use these samples as the extreme sequence set. Each extreme sequence is sampled with three reasoning traces to ensure stable anchors. For the large-scale attack evaluations, one reasoning trace per query sequence is sufficient to reproduce the main results of Blackspectrum; we present the averaged results by sampling the reasoning trace of each sequence three times and report to reduce results variance. For baselines, running DE-COP on modern LRMs incurs prohibitively high API query costs. In addition, many advanced LRMs reject such queries because of copyright safeguards. We therefore include results on two relatively low-cost LRMs on arXivReasoning. (See more details in Appendix~\ref{app: imdetails})

\vspace{-0.5em}
\subsection{Main Results}
Our attack results are evaluated at two levels: sequence-level and document-level. The document-level results are derived from the sequence-level evaluations by aggregating sequences' membership scores~\cite{duan2024membershipinferenceattackswork} in each document. Table~\ref{tab:sentenceperformance} shows that the \textbf{attacks are markedly easier when reasoning traces are exposed}, and even our naive reasoning-trace-based attacks outperform prior methods that rely on repeated queries to probe membership signals without access to reasoning traces. Specifically, once reasoning traces are exposed, substantial attack performance gains of around 23.8\% Balanced ACC, 29.9\% AUC, and 1.9$\times$ TPR@5\%FPR can be achieved. The BlackSpectrum framework achieves consistently strong and stable performance with minimal query cost.  
Some methods designed for traditional pre-training LLMs exhibit unstable performance on LRMs. For example, CDD~\cite{dong-etal-2024-generalization} sometimes even yields inconsistent or counterintuitive results that may contradict the original assumptions. One possible reason is that the post-training alignment in LRMs constrains the original next-token prediction paradigm~\cite{kirk2024understandingeffectsrlhfllm}, so even seen sequences may not yield faithful continuations. 



%

\subsection{Ablations and Discussions}

\subsubsection{Ablation Studies} To assess the impact of each design choice in BlackSpectrum, we ablate three core components on arXivReasoning targeting Gemini-2.5-flash. For the trace encoder model, we evaluate (i) w/o the sequence-specific denoising module and (ii) different pre-trained embedding models on overall framework performance. For the axis builder, we evaluate (iii) w/o the uncertainty-based selection module. Table~\ref{tab:ablation} shows that all components contribute to BlackSpectrum’s effectiveness. 
\subsubsection{Discussions} 
In this section, we study several factors that affect attack effectiveness.

 \begin{figure}[t]
    \centering
    \subfloat[\small AUC w.r.t. \#Seq. Length]{
        \includegraphics[width=0.45\linewidth]{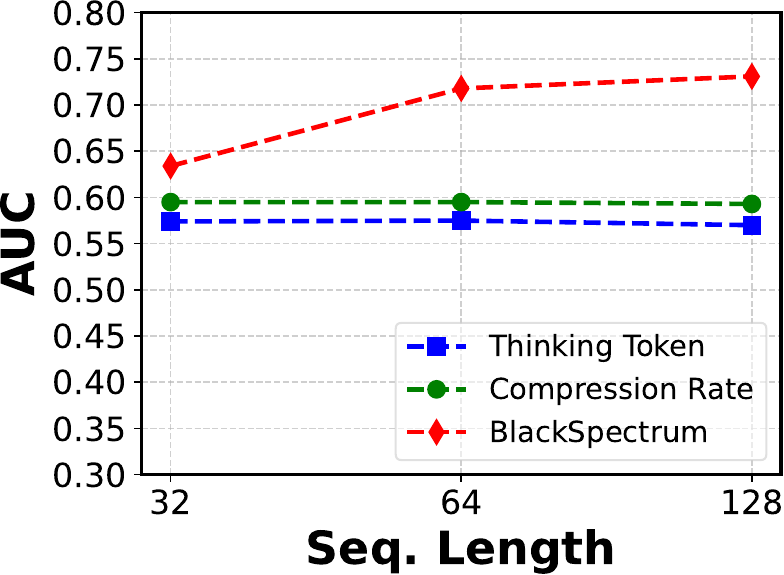}
    }
    \hfill
    \subfloat[\small T@5\%F w.r.t. \#Seq. Length]{
        \includegraphics[width=0.45\linewidth]{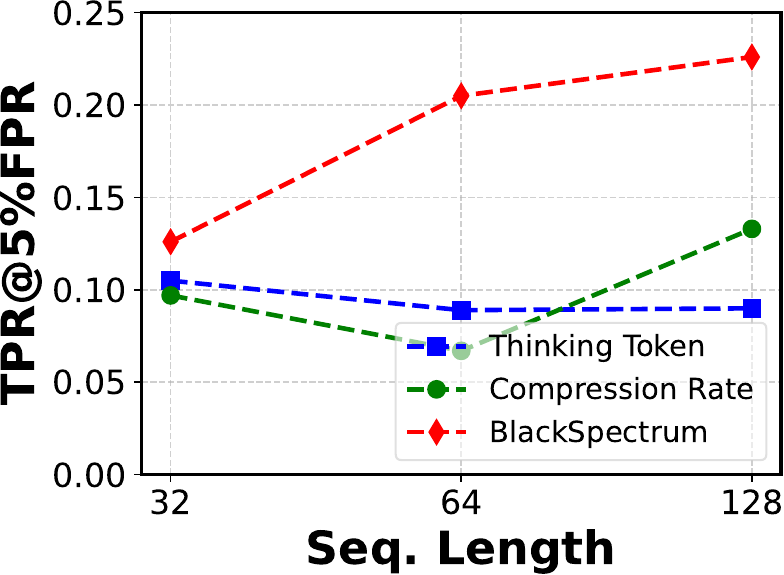}
    }
    \vspace{-0.5em}
    \caption{\small Attack performance w.r.t  Sequence Length.}
    \label{fig: sequencelen}
        \Description{sequencelen}
    \vspace{-0.5em}

\end{figure}

\vspace{0.2em}
\noindent\textbf{Impact of target sequence length}. 
As shown in Figure~\ref{fig: sequencelen}, attack performance generally improves with longer target sequences. Longer sequences carry more information, enabling the reasoning model to better recall seen member texts and distinguish them from non-members. This aligns with prior conclusions on membership inference attacks for pretrained LLMs~\cite{shi2024detecting,kaneko2024samplingbasedpseudolikelihoodmembershipinference,carlini2021extracting,chen-etal-2025-statistical}. 

 \begin{figure}[t]
    \centering
    \subfloat[\small Gemini-2.5-flash]{
        \includegraphics[width=0.46\linewidth]{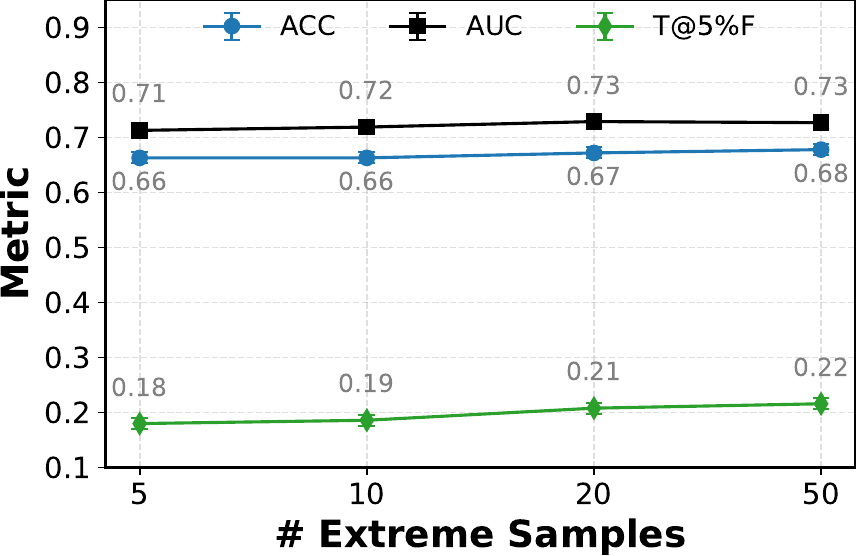}
        \label{fig:extreme-gemini}
         \Description{fig:extreme-gemini}
    }
    \hfill
    \subfloat[\small Claude-sonnet-4]{
        \includegraphics[width=0.46\linewidth]{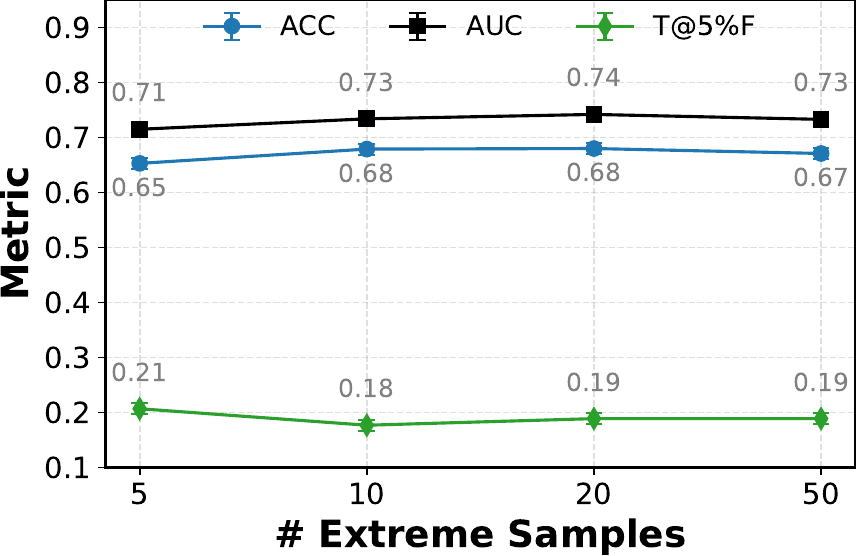}
        \label{fig:extreme-claude}
         \Description{fig:extreme-claude}
    }
    \vspace{-0.5em}
    \caption{\small Attack Performance w.r.t. \#Extreme Sequences}
    \label{fig: numberextreme}
        \Description{number of extreme sequence}
\end{figure}

\vspace{0.2em}
\noindent\textbf{Impact of the number of extreme sequences}. We examine how the number of extreme samples affects attack performance. Figure~\ref{fig: numberextreme} shows that only 5 verbatim recall and 5 synthetic sequences are sufficient to achieve strong attack performance. Increasing the number of sequences yields slight performance gains. This suggests that the key is to obtain stable and high-quality anchors rather than simply relying on sample quantity.

 \begin{figure}[t]
    \centering
    \subfloat[\small arXivReasoning]{
        \includegraphics[width=0.46\linewidth]{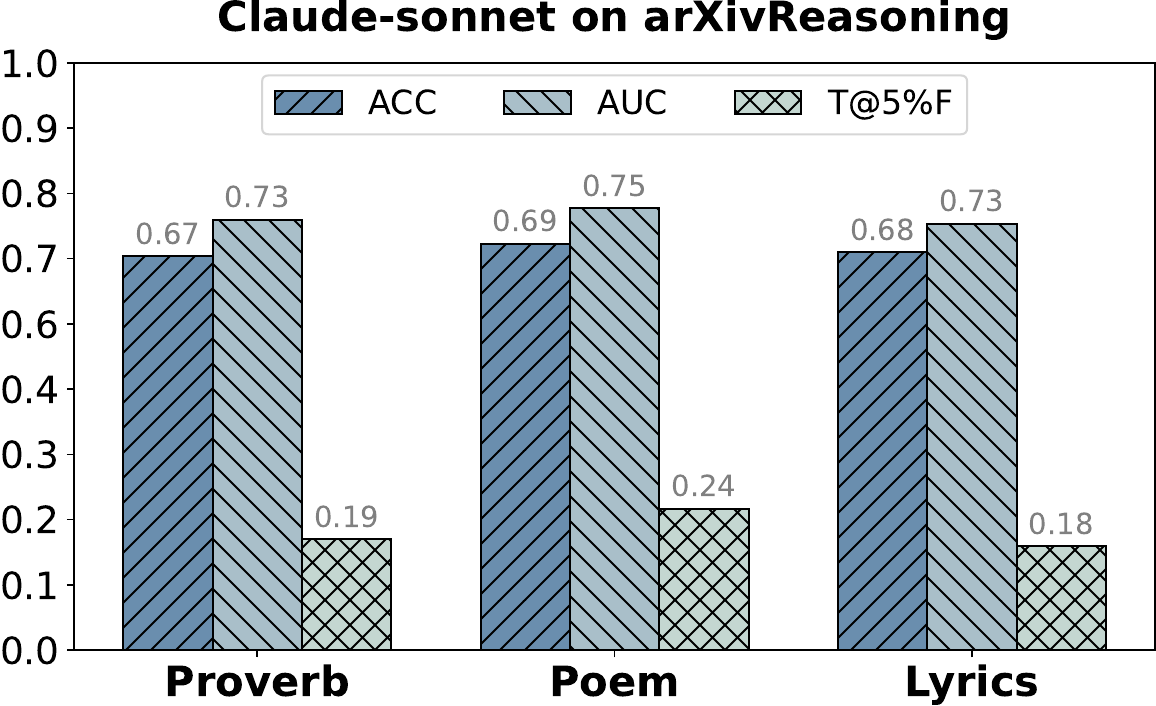}
    }
    \hfill
    \subfloat[\small BookReasoning]{
        \includegraphics[width=0.46\linewidth]{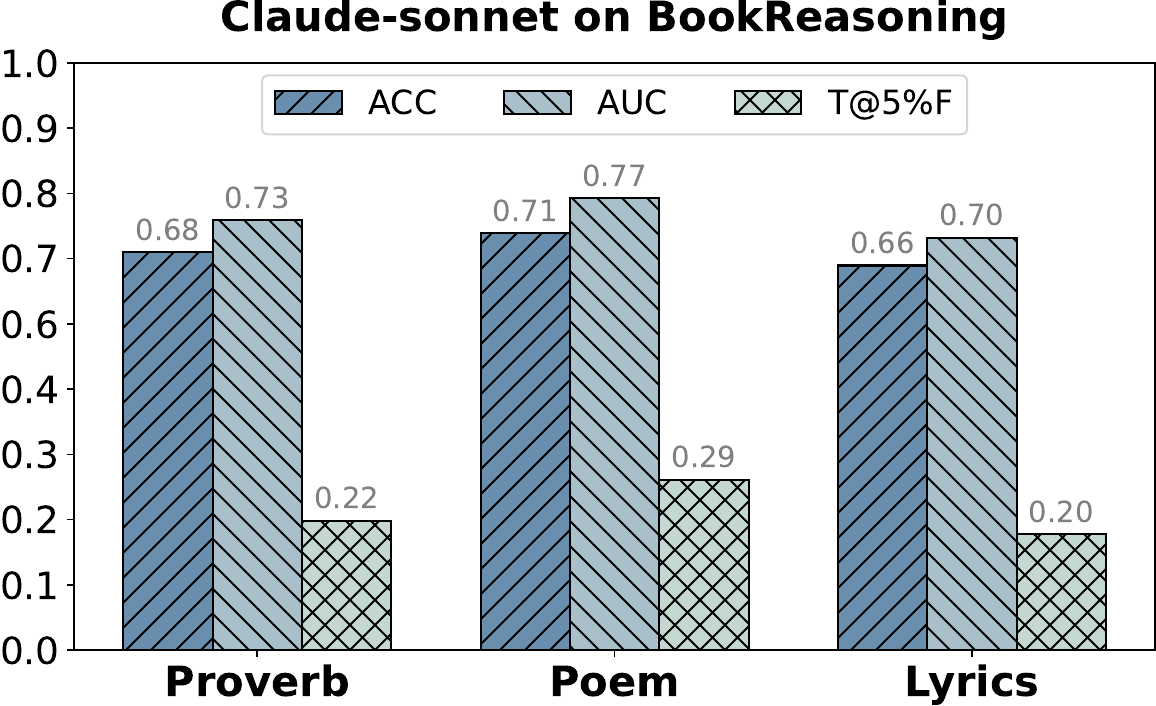}
    }
    \vspace{-0.5em}
    \caption{\small Attack Performance w.r.t. Extreme Sequence Type}
    \label{fig: typeextreme}
        \Description{extremesequencetype}
    \vspace{-0.5em}
\end{figure}

\vspace{0.2em}
\noindent\textbf{Impact of the type of extreme sequences}.
We further analyze the impact of the extreme sequence types on attack performance. Figure~\ref{fig: typeextreme} shows that proverbs, poems, and lyrics achieve comparable results, with poems yielding marginally higher AUC and TPR@5\%FPR. This suggests that the specific content type is less critical than its ability to capture stable recall–inference reasoning modes, offering flexibility in anchor selection for practical attacks.

\subsubsection{Reasoning traces versus direct output answers on the membership signal.}


To examine whether reasoning traces leak more membership signals than final answers, we adapt BlackSpectrum into two variants: (i) Answer-only, which encodes only final answers; and (ii) Concatenation, which combines reasoning traces with answers. As shown in Table~\ref{tab:output-only}, Answer-only performs notably worse, suggesting that answers alone reveal limited membership features, while Concatenation slightly underperforms Trace-only, implying that adding answers may introduce noise.

\begin{table}[t]
\centering
\renewcommand{\arraystretch}{0.8}
\vspace{\baselineskip}
\caption{Intermediate reasoning traces vs. direct output answers targeting Claude-sonnet-4. Final answers contain limited membership signals.}
\large
\label{tab:output-only}
\resizebox{0.95\columnwidth}{!}{%
\begin{tabular}{l|l|ccc}
\specialrule{1.1pt}{0pt}{0pt}
\textbf{Dataset} & \textbf{Variants} & \textbf{ACC} & \textbf{AUC} & \textbf{T@5\%F} \\ 
\midrule
\multirow{3}{*}{arXivReasoning} 
 & Answer-only     &0.510\textcolor{gray}{$\downarrow$}& 0.523\textcolor{gray}{$\downarrow$}& 0.087\textcolor{gray}{$\downarrow$}\\
 & Concatenation   & 0.728\; & 0.668\;& 0.196\; \\
 & Trace-only (original) & 0.733\;  & 0.671\;  & 0.189\;  \\
\midrule
\multirow{3}{*}{BookReasoning} 
 & Answer-only     & 0.589\textcolor{gray}{$\downarrow$} &0.568\textcolor{gray}{$\downarrow$} & 0.119\textcolor{gray}{$\downarrow$} \\
 & Concatenation   & 0.718\; & 0.668\; & 0.199\; \\
 & Trace-only (original) & 0.732\; & 0.678\; & 0.220\; \\
\specialrule{1.1pt}{0pt}{0pt}
\end{tabular}
}

\end{table}

\begin{table}[t]
\vspace{-0.5em}

\centering
\renewcommand{\arraystretch}{0.8}
\vspace{\baselineskip}
\caption{Effect of Reasoning Detail Level on Membership Leakage. High-level compressed reasoning summaries lead to weaker membership signals.}
\large
\label{tab: case study}
\resizebox{0.95\columnwidth}{!}{%
\begin{tabular}{l|ccccc}
\specialrule{1.1pt}{0pt}{0pt}
\textbf{Variants} & \textbf{ACC} & \textbf{AUC} & \textbf{T@5\%F} & \textbf{P-value}$\downarrow$ & \textbf{ES}$\uparrow$ \\ 
\midrule
Original    & 0.746\;& 0.705\;&0.330\;&$\text{1.523} \times 10^{-9}$  & 0.897 \\
\midrule
Mild compression    &0.685\textcolor{green}{$\downarrow$}&0.676\textcolor{green}{$\downarrow$}& 0.050\textcolor{green}{$\downarrow$} & $\text{1.317} \times 10^{-5}$  & 0.633 \\
Strong compression  &0.565\textcolor{green}{$\Downarrow$} & 0.552\textcolor{green}{$\Downarrow$}& 0.030\textcolor{green}{$\Downarrow$}&$\text{3.363} \times 10^{-1}$  & 0.129 \\
\specialrule{1.1pt}{0pt}{0pt}
\end{tabular}
}
\vspace{-0.5em}
\end{table}

\subsubsection{Case Study: Effect of Reasoning Detail Level on Membership Leakage}
In this section, we perform a case study to examine how the level of reasoning detail affects membership leakage and the trade-off between interpretability and privacy.
We introduce a compression module that rewrites each trace into a concise summary, reducing fine-grained details while preserving the core reasoning logic. The compressed traces are then evaluated under the same attack pipeline for comparison with the original (uncompressed) setting.
For this case study, we use the same 200 samples used in our preliminary exploration section. The target LRM is Claude-sonnet-4. For the compression module, we prompt \textit{GPT-3.5-turbo} to compress reasoning traces into concise, high-level text while preserving key logic. We examine two compression levels: compressing into 5\textasciitilde6 sentences (mild) and 2\textasciitilde3 sentences (strong). The detailed example is provided in Appendix~\ref{app: casedetails}. As shown in Table~\ref{tab: case study}, we can observe that compressing reasoning traces weakens the attack performance. This indicates that concise summaries can mitigate membership leakage. Our findings highlight the importance for LRM companies to carefully balance transparency and privacy when exposing reasoning traces via APIs. 


\section{Conclusion}

In this work, we initiate the systematic exploration of MIAs on black-box LRMs. Our analysis shows that intermediate reasoning traces contain strong membership signals,  forming a continuous recall–inference spectrum phenomenon in the latent space. Based on this observation, we propose BlackSpectrum, a novel MIA framework that leverages intermediate reasoning traces to infer membership in modern black-box LRMs. Moreover, we contribute two new datasets, arXivReasoning and BookReasoning, and five reasoning-trace-based baselines to support future MIA research on modern LRMs. In future work, we will explore finer-grained reasoning-trace analysis to identify which reasoning steps contribute most to membership leakage and extend the evaluation to a broader range of commercial LRMs.
%
\bibliographystyle{ACM-Reference-Format}
\bibliography{sample-base}

\appendix
\section{Method Supplementary Details}\label{app: methoddetails}
\begin{figure}[t]
  \centering  
  \includegraphics[width=1\columnwidth]{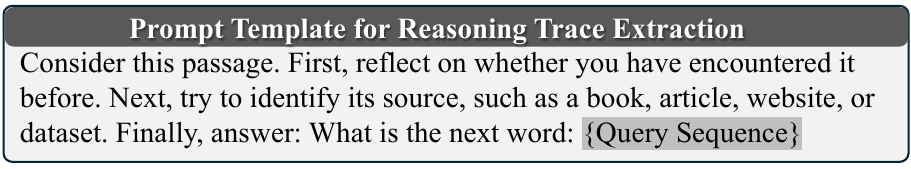}
  \caption{Prompt template for reasoning trace extraction.
 }
  \label{fig: prompt}
        \Description{app}
   \vspace{-0.5em}
\end{figure}

This section presents the template employed for reasoning trace extraction. The design of this template adheres to two essential principles. The prompt template is illustrated in Figure~\ref{fig: prompt}. The overall workflow of our method is summarized in Algorithm~\ref{alg: blackspectrum}.

\section{Experimental Details}

\subsection{Naive Attacks on Black-box LRMs}\label{sec: naive}

In this section, we provided details about our designed naive attacks that rely on surface-level cues extracted from the reasoning traces of black-box LRMs.

\subsubsection{Count of thinking tokens} Prior work~\cite{muennighoff2025s1simpletesttimescaling,qian2025demystifying} has observed that certain indicative tokens (\emph{e.g.}, ``Hmm,'' ``Wait,'' or ``so''), referred to as thinking tokens, tend to emerge during the reasoning process and often signify that the LRM is converging toward a correct answer. Inspired by this observation, we hypothesize that a higher count of thinking tokens in reasoning traces indicates queries derived from non-member sequences, whereas fewer thinking tokens suggest member sequences. This is because, for unfamiliar sequences, a higher count of thinking tokens reflects the LRM's need for multiple attempts and adjustments to approach the answer.

\begin{figure}[t]
  \centering  
  \includegraphics[width=0.98\columnwidth]{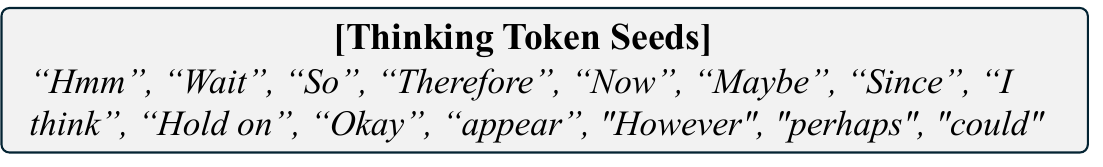}
  \caption{Thinking token seeds.
 }
  \label{fig: seed}
   \vspace{-0.5em}
\end{figure}


Technically, as shown in Figure~\ref{fig: seed}, we construct a set of thinking token seeds $\mathcal{T} = \{t_1, t_2, \dots, t_M\}$ that frequently appear in reasoning traces~\cite{muennighoff2025s1simpletesttimescaling,qian2025demystifying}. 
For each reasoning trace $r$, the text is segmented into words and $n$-grams (up to trigrams) after stopword removal, forming a candidate set of possible thinking phrases $\mathcal{C}(r) = \{c_1, c_2, \dots, c_N\}$.
We then compute the semantic similarity $\text{sim}(c_i, t_j)$ between each candidate $c_i$ and each seed $t_j$, and select candidates whose maximum similarity exceeds a predefined threshold $\delta$ (set to $0.8$ in our experiments):
\[
\mathcal{T}(r) = \{\, c_i \in \mathcal{C}(r) \mid \max_{t_j \in \mathcal{T}} \text{sim}(c_i, t_j) \ge \delta \,\}.
\]
The total number of such tokens constitutes the \textit{thinking token count}, which serves as the membership score:
\[
\epsilon_{\text{think}}(r) = |\mathcal{T}(r)|.
\]
A higher $s_{\text{think}}(r)$ indicates more exploratory or uncertain reasoning, suggesting that the query sequence $s$ is more likely to be a non-member.
\begin{figure}[t]
  \centering  
  \includegraphics[width=0.9\columnwidth]{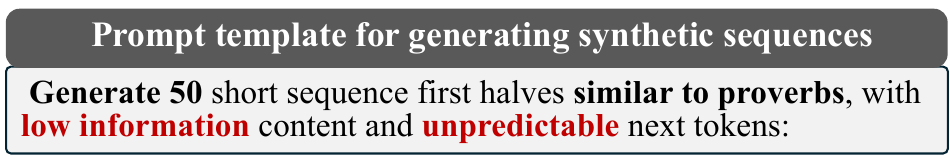}
  \caption{Prompt template for generating synthetic sequences.The same generation template as used in the method section is applied here.
 }
  \label{fig: prompt-generate}
      \Description{app}
   \vspace{-0.5em}
\end{figure}

\begin{figure}[t]
  \centering  
  \includegraphics[width=1\columnwidth]{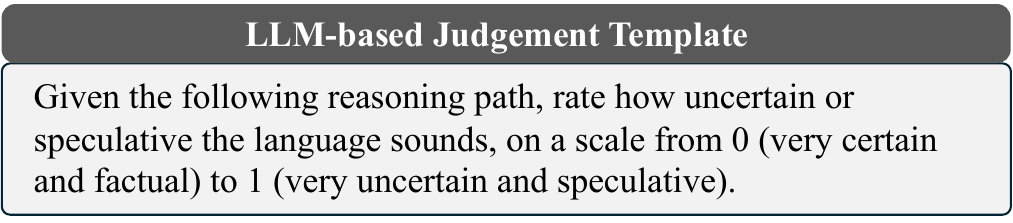}
  \caption{LLM-based judgement Template
.
 }
  \label{fig: judge}
        \Description{app}
   \vspace{-0.5em}
\end{figure}

\subsubsection{Compression Rate}

The compression rate of a reasoning trace reflects how much information it contains. For unfamiliar sequences, the traces tend to involve parallel options, hesitation, and exploratory reasoning, which makes them richer in information and thus more difficult to compress. 
%

Techniqually, Following the information-theoretic formulation of~\citet{grunwald2004shannoninformationkolmogorovcomplexity} and~\citet{morris2025languagemodelsmemorize}, 
we approximate the compression rate using the per-token \textit{Negative Log-Likelihood (NLL)} under a reference language model (e.g., GPT-2~\cite{radford2019language}):

\[
\epsilon_{\text{comp}}(r) = -\frac{1}{T} \sum_{t=1}^{T} \log p_{\text{ref}}(x_t \mid x_{<t}),
\]

where $r$ denotes the reasoning trace, $x_{1:T}$ is its token sequence of length $T$, and $p_{\text{ref}}$ represents the conditional probability distribution given by the reference model (\emph{e.g.}, GPT-2~\cite{radford2019language}).
A higher $\epsilon_{\text{comp}}(r)$ indicates lower compressibility and hence greater informational richness, implying that the reasoning process is less stereotyped or more exploratory, and that the corresponding input is more likely to be a non-member sample.x


\subsubsection{Trace Consistency}

Inspired by prior work~\cite{consistency,dong-etal-2024-generalization}, we measure the \textit{consistency} of reasoning traces as an indicator of the model’s familiarity with specific content. 
Specifically, for each query $q$, we collect $K$ reasoning traces $\{r_1, r_2, \dots, r_K\}$ generated by repeated sampling. 
The trace consistency is then quantified as the average pairwise edit distance (character-level~\cite{levenshtein1966binary} or token-level~\cite{dong-etal-2024-generalization}) among these traces:

\[
\epsilon_{\text{cons}}(s) = \frac{2}{K(K-1)} \sum_{1 \le i < j \le K} d_{\text{edit}}(r_i, r_j),
\]

where $d_{\text{edit}}(\cdot,\cdot)$ denotes the normalized edit distance between two reasoning traces.
A lower $\epsilon_{\text{cons}}(s)$ indicates higher response stability and stronger familiarity with the sequence (member), 
while a higher value suggests variability or uncertainty in reasoning, implying that the input query sequence $s$ is more likely to be a non-member.

%

\subsubsection{LLM-based judgement} Intelligent LLMs can be employed to assess text sentiment polarity or data contamination~\cite{yang2023rethinkingbenchmarkcontaminationlanguage}. In our setting, we similarly leverage a third-party LLM (\emph{e.g.}, GPT-3.5-turbo) to assign an uncertainty score to each reasoning trace (See prompt in Figure~\ref{fig: judge}). A higher score reflects greater certainty, suggesting that the query is derived from member samples.

\begin{algorithm}
\caption{BlackSpectrum}\label{alg: blackspectrum}
\begin{algorithmic}[1]
\Require Target LRM $\Theta$, query sequence $s_i$, encoder $E(\cdot)$, 
recall-like anchor $\mathbf{a}_{\mathcal{M}}$, inference-like anchor $\mathbf{a}_{\mathcal{N}}$, threshold $\lambda$
\Ensure Membership decision for $s_i$

\State \textbf{Step 1: Reasoning Trace Extraction.}
Feed a prompt combining the designed instruction and $s_i$ into $\Theta$ to obtain the reasoning trace $r_i$ that reflects the model’s internal reasoning process.

\State \textbf{Step 2: Encoding and Denoising.}
Encode the reasoning trace and sequence representations:
\[
\mathbf{e}_{r_i}=E(r_i), \quad
\mathbf{e}_{s_i}=E(s_i).
\]
Remove the sequence-specific component to obtain a denoised reasoning embedding:
\[
\tilde{\mathbf{e}}_{r_i}
= \mathbf{e}_{r_i} -
\frac{\langle \mathbf{e}_{r_i}, \mathbf{e}_{s_i}\rangle}
{\|\mathbf{e}_{s_i}\|^2}\mathbf{e}_{s_i}.
\]

\State \textbf{Step 3: Projection-based Membership Prediction.}
Compute the axis direction 
$\mathbf{\hat{u}}=\frac{\mathbf{a}_{\mathcal{N}}-\mathbf{a}_{\mathcal{M}}}{\|\mathbf{a}_{\mathcal{N}}-\mathbf{a}_{\mathcal{M}}\|}$ 
and distance $D=\|\mathbf{a}_{\mathcal{N}}-\mathbf{a}_{\mathcal{M}}\|$. 
Project the denoised embedding:
\[
\rho_{s_i}=(\tilde{\mathbf{e}}_{r_i}-\mathbf{a}_{\mathcal{M}})^{\top}\hat{\mathbf{u}},
\]
then normalize to obtain the membership score:
\[
\epsilon_{s_i}=1-\frac{\rho_{s_i}}{D}.
\]
Return \textbf{member} if $\epsilon_{s_i}\ge\lambda$, otherwise \textbf{non-member}.

\Statex
\textit{Note: Since AUC is adopted as the evaluation metric, no fixed threshold is required; MIA performance is evaluated under varying $\lambda$ values, following prior work~\cite{zhang2025mink, shi2024detecting}.}
\end{algorithmic}
\end{algorithm}

\begin{table*}[t]
   \caption{The statistics of arXivReasoning and BookReasoning datasets.}
   \centering
   \renewcommand{\arraystretch}{1} 
   \resizebox{0.98\textwidth}{!}{
   \begin{tabular}{lccccc}
       \toprule
       \textbf{Datasets} & \textbf{Type} & \textbf{Sequence Len.} & \textbf{\# Seq. (per Len.)} & 
       \textbf{Resources} & \textbf{Target Model}
        \\
               \midrule
              \multirow{2}{*}{arXivReasoning} 
         & Member     & 32/64/128 & 761  & 25 arXiv papers serve as members in~\cite{duarte2024decop,shi2024detecting} & \multirow{2}{*}{\makecell{LRMs whose knowledge \\ cutoff predates August 2025}} \\ 
         & Non-member & 32/64/128 & 844  & 30 papers released after May 2025 &  \\
       \midrule
       \multirow{2}{*}{BookReasoning} 
         & Member     & 32/64/128 & 3505 & 105 bestselling books serve as members in~\cite{duarte2024decop,shi2024detecting,cooper2025extractingmemorizedpiecescopyrighted,zhang2025pretrainingdatadetectionlarge} & \multirow{2}{*}{\makecell{LRMs whose knowledge \\ cutoff predates May 2025}} \\ 
         & Non-member & 32/64/128 & 2606 &\makecell{79 books published after May 2025}&  \\

       \bottomrule
   \end{tabular}}
   \label{tab: detaileddataset}
\end{table*}

\subsection{Baselines and Implementation details} \label{app: imdetails}

In this section, we provide additional implementation details for the baseline methods.  For the SaMIA baseline applied to conventional pre-training LLMs, we follow the original implementation: each sentence is truncated to its first half as a prefix, and the model is tasked to generate the remaining half as a suffix.
The generated suffix is then compared with the ground-truth continuation using ROUGE scores~\cite{lin2004rouge}.
This setup effectively evaluates the model's ability to perform autoregressive completion, either through native completion APIs (\emph{i.e.,}\texttt{completion}) or by simulating the closed-source completion behavior with open-source models. 

However, for modern LRMs, such as Gemini-2.5, Claude-Sonnet, and GPT-5-mini, direct completion interfaces (\emph{i.e.,}\texttt{completion}) are not always available. In these cases, we instead design instruction-based prompts that explicitly ask the model to continue the given context (\emph{i.e.,}\texttt{chat.completion}).
Although this approach inevitably produces \textit{intermediate reasoning traces} interleaved with the output continuation, we strive to make it as consistent as posysible with the original SaMIA evaluation setting.  Specifically, we use \texttt{\{"role": "system", "content": "You are a helpful assistant. Please answer directly, do not think."role": "user", "content": "Generate a continuous sentence of 10 tokens? Please direct answer"\}}. This instruction helps constrain our target LRM to produce direct continuations, reducing the occurrence of lengthy intermediate reasoning traces.
This also suggests that \textbf{prior attack schemes of this kind are not fully compatible with modern LRMs}.
For the CDD~\cite{dong-etal-2024-generalization}baseline, we adopt a similar procedure by generating multiple continuous suffixes and measuring the token-level edit distance among the generated suffixes themselves. 

For baselines, running DE-COP on modern LRMs incurs prohibitively high API query costs.
This is primarily because LRMs are substantially more expensive and slower to query than conventional pre-training LLMs.  In addition, many advanced LRMs reject such queries because of copyright safeguards. For example, \textit{``I cannot determine which passage is verbatim from the Martha Wells work without... memorizing copyrighted text or reproducing it for comparison purposes, which I should avoid...''}. We therefore include results on two relatively low-cost LRMs evaluated on the arXivReasoning benchmark, which do not trigger copyright safeguard refusals. This also highlights that, for black-box LRM attacks, reducing API costs and preventing over-thinking (or bypassing safeguards) are important directions for future research.


\subsection{Case Study: Effect of Reasoning Detail Level on Membership Leakage}\label{app: casedetails}

\begin{figure}[t]
  \centering  
  \includegraphics[width=0.8\columnwidth]{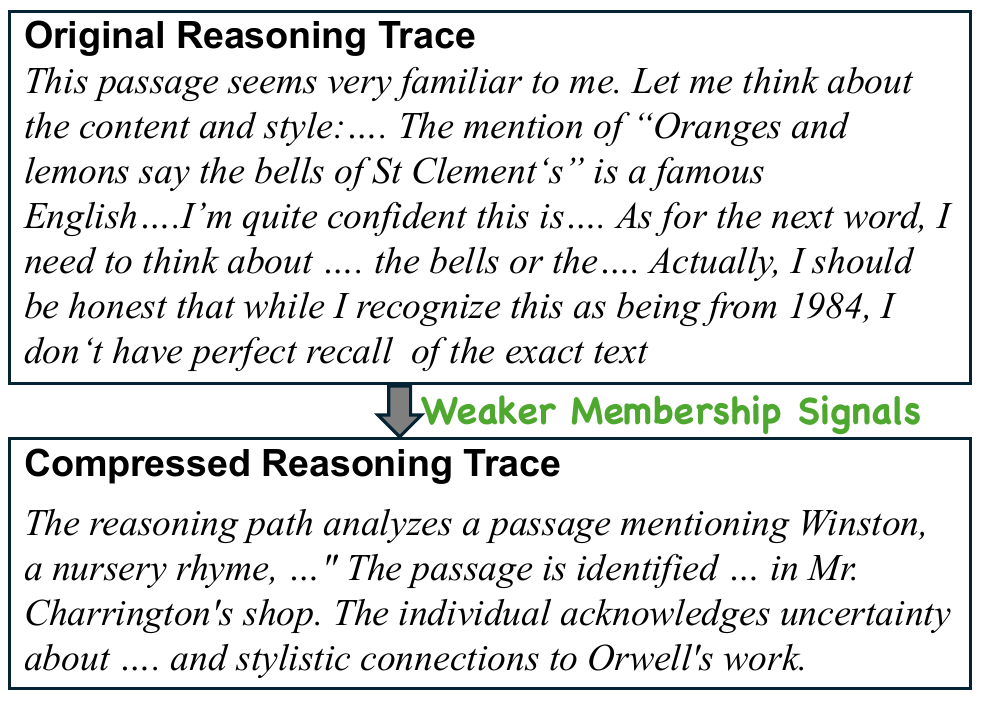}
  \caption{A real-world example on Claude-sonnet-4. It shows that compressing detailed reasoning traces weakens membership signals: The compression of reasoning traces eliminates fine-grained recall mode, transforming explicit memory recall into abstract logical statements.
 }
  \label{fig: compress}
        \Description{app}
   \vspace{-0.5em}
\end{figure}

In this section, we provide the detailed example corresponding to Figure~\ref{fig: compress}, as well as the prompt used to guide GPT-3.5-turbo for reasoning trace compression.
The prompt is shown below: \texttt{\{"role": "system", "content": "You are a helpful assistant who summarizes reasoning paths into concise summaries.", "role": "user", "content": Please read the following reasoning path and provide a concise summary in 5-6 sentences.\}}

\section{Dataset Details}\label{app: datasetdetails}

\paragraph{Limitations of Existing Datasets for MIAs}Existing MIA datasets, originally developed for pre-training language models, are no longer suitable for modern LRMs.
Many of the materials contained in these datasets may have already been included in the training corpora of recent LRMs, undermining their validity as evaluation benchmarks.
In particular, they were designed based on a knowledge cutoff rule, where labeled non-member samples are defined as those published after the pretraining language model's knowledge cutoff date. For recent emerging LRMs, the member–non-member labeling is unreliable. As their training corpus is updated, some materials labeled as non-members may already be included. 
Some prior gray-box attacks used the MIMIR dataset~\cite{duan2024membershipinferenceattackswork} built on the Pile corpus with smaller pre-training LLMs; however, no LRMs currently match this training corpus, so we do not adopt it in this work.

\paragraph{Data Construction and Statistics}
To provide a reliable benchmark tailored for LRMs, we construct two new reasoning-oriented datasets, arXivReasoning and BookReasoning. The dataset statistics are summarized in Table~\ref{tab: detaileddataset}. Specifically:

\begin{itemize}

    \item \textit{arXivReasoning.} Members are selected from verified papers identified by~\cite{duarte2024decop}, while non-members include 55 papers released on arXiv after May 2025, ensuring our target LRM has never seen these materials. The sequences are similarly divided into 32, 64, and 128 tokens, yielding 761 member sequences and 844 non-member sequences in total.
    
    \item   \textit{BookReasoning.} Members are sourced from verified copyrighted books identified by ~\cite{shi2024detecting, duarte2024decop, zhang2024pretrainingdatadetectionlarge, chang-etal-2023-speak}, which confirms inclusion in the training data of most language models. Non-members consist of 79 books published after May 2025, ensuring they remain unseen by LRMs whose knowledge cutoff predates that date. Each sequence is segmented into lengths of 32, 64, and 128 tokens, resulting in 3,505 member sequences and 2,606 non-member sequences across all lengths.
\end{itemize}


\end{document}